\documentclass[%
 reprint,
superscriptaddress,
nofootinbib,
 amsmath,amssymb,
 aps,
]{revtex4-2}

\usepackage{graphicx}
\usepackage{dcolumn}
\usepackage{bm}
\usepackage{xcolor}
\usepackage{hyperref}

\begin{document}

\preprint{APS/123-QED}

\title{On the cosmological evolution of Scalar Field Dark Matter in the \texttt{CLASS} code: accuracy and precision of numerical solutions}


\author{L. Arturo Ure\~na-L\'opez}
 \homepage{lurena@ugto.mx}
\affiliation{%
 Departamento de F\'isica, DCI, Campus Le\'on, Universidad de Guanajuato, 37150, Le\'on, Guanajuato, M\'exico
}%

\author{Francisco X. Linares Cede\~no}
 \homepage{francisco.linares@umich.mx}
\affiliation{
 Instituto de F\'isica y Matem\'aticas, Universidad Michoacana de San Nicol\'as de Hidalgo, Edificio C-3, Ciudad Universitaria, CP. 58040 Morelia, Michoac\'an, M\'exico.
}%


\date{\today}

\begin{abstract}
We present a numerical analysis of the cosmological evolution of scalar field dark matter (SFDM) in the Boltzmann code \texttt{CLASS}, based on a dynamical system analysis of previous works. We show a detailed study of the evolution of the different dynamical variables, and in particular of the energy density and its corresponding linear perturbations. The numerical results are in good agreement with those of the original SFDM equations of motion, and have better accuracy than other approaches. In addition, we calculate the temperature and matter power spectra and discuss the reliability of their numerical results. We also give simple examples in which we can put constraints on the field mass using recent likelihoods incorporated in the Monte Carlo Markov Chain sampler \texttt{MontePython}.
\end{abstract}

\maketitle


\section{\label{sec:level1}Introduction}

In the modern era of cosmology~\cite{Planck:2018vyg,Peebles:2022akh}, it is mandatory to develop theoretical models able to describe the universe at large scales, with the precision that current data demand~\cite{Peebles:2022akh}. One key theoretical ingredient, particularly important for the formation of cosmic structure, is dark matter (DM), which must capture the physics of such a process in a mathematically consistent way~\cite{RevModPhys.90.045002}. This always starts, for the cosmological setting, with properly solving the Einstein-Boltzmann system describing the cosmological evolution for the initial perturbations of both the matter components and the metric tensor at the linear level~\cite{Ma:1995ey}. 

Different Boltzmann solvers have been programmed for the linearized form of the Einstein equations in a Friedmann-Robertson-Walker-Lema\^itre (FRWL) universe, such as \texttt{CMBFAST}~\cite{Seljak:1996is}, \texttt{CMBEASY}~\cite{Doran:2003sy}, \texttt{CAMB}~\cite{Lewis:1999bs} and \texttt{CLASS}~\cite{Lesgourgues:2011re}. Only the latter two, \texttt{CAMB} and \texttt{CLASS}, have been kept up to date, and both are used by the community of cosmologists. The aforementioned codes consider the cold dark matter (CDM) model as the main matter component, and they are very well suited to explore most of its properties, although they have been amended to study alternative dark matter models in recent years.

One of these alternative models to CDM, which is currently one of the compelling proposals that has been explored for the last two decades, is that based on a scalar field. It is found in the literature under several names: Scalar Field Dark Matter (SFDM), Ultralight Axions, Fuzzy Dark Matter, Axion-like particle, Bose-Einstein Condansate, Wave Dark Matter (some initial works on this model are~\cite{Abbott:1982af,Matos:1998du,Matos:1998vk,Matos:2000ng,hu2000fuzzy,Hu:2000ke,Masso:2002ip}, whereas more recent work can be found at~\cite{Urena-Lopez:2015gur,Hlozek:2014lca,Hui:2016ltb,Irsic:2017yje,Urena-Lopez:2019kud,Kulkarni:2020pnb,Ferreira:2020fam,Foidl:2022bpn,Hartman:2022cfh}). All these names reflect the particular properties of this DM particle. Its mathematical description is given by a scalar field $\phi$ (which can be real or complex), with a fiducial ultralight mass of $m_{\phi} c^2 \sim 10^{-22}$eV that can be produced by the Peccei-Quinn mechanism for pseudo-Goldstone bosons (as is the case for QCD axions) and whose quantum nature manifests itself at cosmological scales through its imprint on the structure formation at small scales. Therefore, we shall refer to this model as SFDM hereafter.

In order to solve the cosmological dynamics of the linear perturbations of SFDM, \texttt{CAMB} have been modified to include a real scalar field as a DM component. This version has been called \texttt{AxionCAMB}~\cite{Hlozek:2014lca}, and it is written for a scalar field endowed with a quadratic potential $V(\phi) \propto \phi^2$. This is why this case is called the \textit{ free case}. On the other hand, \texttt{CLASS} has been modified by the authors of this work with the aim of including the SFDM model with quadratic potential: \texttt{class.FreeSF}~\cite{Urena-Lopez:2015gur}. Both codes deal with the background dynamics and linear perturbations of the SFDM, and it is possible to obtain the anisotropies of the Cosmic Microwave Background radiation (CMB), as well as the Matter Power Spectrum (MPS) for several mass values of the SFDM particle~\footnote{Repositories for \texttt{AxionCAMB} and \texttt{class.FreeSF} based on earlier versions of \texttt{CAMB} and \texttt{CLASS} respectively, can be found at \url{https://github.com/dgrin1/axionCAMB} and \url{https://github.com/lurena-lopez/class.FreeSF}.}.

Whereas the only SFDM model that has been treated with \texttt{AxionCAMB} is that of the noninteracting case, the authors made new amendments to \texttt{CLASS} to study potentials with self-interaction in the scalar field potential. For example, the full axion-like potential $V(\phi) \propto \cos(\phi)$ was fully implemented for the first time in \texttt{CLASS} in the work~\cite{Cedeno:2017sou}, where it was shown that this model presents an excess of power on small scales in the MPS with respect to CDM~\footnote{Recently, this kind of features in the MPS have also been reported for a scalar field with a non-periodic potential~\cite{Chatrchyan:2023cmz}.}, with a greater discussion of the cosmological signatures of this SFDM model in~\cite{LinaresCedeno:2020dte}. The other SFDM potential we have considered in \texttt{CLASS} is that of a hyperbolic function $V(\phi)\propto \cosh(\phi)$~\cite{Urena-Lopez:2019xri}, where the self-interacting term has the opposite sign to that of axions. 

As we will show later, the mathematical treatment of the SFDM cosmological evolution used in the modified versions of \texttt{CLASS} makes use of new dynamical variables based on a dynamical system analysis. We have shown that this is useful for a unified description of different scalar potentials, showing our own method to deal with tree different cases in a one-parametric way~\cite{LinaresCedeno:2021sws}.

The modified versions of \texttt{CLASS} mentioned above for SFDM deal with the linear evolution of density perturbations. This information (initial conditions at radiation-dominated era, physical effects on observables such as CMB, MPS, Halo Mass Function and others) turns out to be crucial for the subsequent realization of realistic numerical simulations of structure formation in the nonlinear regime, see for example~\cite{Medellin-Gonzalez:2020eww,Medellin-Gonzalez:2023kxz,Mocz:2023adf,Banares-Hernandez:2023axy,Sipp:2022tdp}.

Our main goal in the present work is to give more details of the mathematical approach we have used in previously amended versions of \texttt{CLASS}, and in turn to show its robustness in describing the physical processes of SFDM up to the level of linear density perturbations. This is done here for SFDM endowed with a quadratic potential,
\begin{equation}
    V(\phi) = \frac{1}{2}m_{\phi}^2 \phi^2 \, , \label{quadratic}
\end{equation}
where $m_\phi$ is the mass of the scalar field particle and the only free parameter in the model. Note that we are using natural units with $h=c=1$, and then $m_\phi$ is given in units of eV. Whenever appropriate, we will compare our method to the common fluid approximation to the SFDM dynamics used in other works, following the description in Refs.~\cite{Passaglia:2022bcr,Cookmeyer:2019rna,Hlozek:2014lca}

This paper is organized as follows. In Sec.~\ref{sec:math-back}, we develop the mathematical formulation for the evolution of the background and linear density perturbations, in terms of new dynamical variables that are appropriate to handle the particularities of SFDM. We also establish the appropriate initial conditions for the scalar field to behave as the DM component at late times, for both background and linear quantities. 

Section~\ref{sec_osc} is dedicated to the description of the typical regime of rapid oscillations of SFDM at late times in its evolution and to the way in which we deal with them for their reliable numerical computation. In particular, we show a detailed study of the evolution of the (barotropic) equation of state, the energy density, and finally the linear density perturbations. As an application of our method in the Boltzmann code \texttt{CLASS}, we present the temperature and matter power spectra and discuss the reliability of our numerical results. Moreover, we also give simple examples in which we can put constraints on the field mass $m_\phi$ using recent likelihoods incorporated in the Monte Carlo Markov Chain (MCMC) sampler \texttt{MontePython}~\cite{Audren:2012wb,Brinckmann:2018cvx}, taking advantage of its close interoperability with \texttt{CLASS}. 

The comparison and equivalence between our approach and the original formulation in terms of the scalar field itself is presented in Sec.~\ref{comparison}. We explicitly show the transformation between our variables and the original field ones $(\phi\, , \dot{\phi})$, and that the numerical results of our method are completely equivalent to the standard field approach. Finally, in Sec.~\ref{discussion} we summarize and discuss our results, highlighting the advantages of our method.

\section{\label{sec:math-back} Mathematical background}
In this section, we present the equations of motion for the SFDM model in the context of an expanding universe, and the subsequent transformations we use to make them more suitable for numerical computations. The original motivations and some extra details of the method described here can be found in~\cite{Urena-Lopez:2015odd,Urena-Lopez:2015gur,Cedeno:2017sou}.

\subsection{\label{sec:back-quant} Background evolution}
Let us consider a spatially-flat FRWL line element,
\begin{equation}
ds^2 = -dt^2 + a^2(t)\left[ dr^2 + r^2\left(d\theta^2 + \sin^2\theta d\phi^2\right) \right]\, ,
\label{flrwsf}
\end{equation}
where $a(t)$ is the scale factor. The background equations for ordinary matter, which is represented by perfect fluids with density $\rho_j$ and pressure $p_j$, as well as for SFDM field $\phi$ endowed with the potential~\eqref{quadratic}, are given by
\begin{subequations}
\label{eq:2}
  \begin{eqnarray}
    H^2 &=& \frac{\kappa^2}{3} \left( \sum_j \rho_j +
      \rho_\phi \right) \, , \label{eq:2a} \\
    \dot{H} &=& - \frac{\kappa^2}{2} \left[ \sum_j (\rho_j +
      p_j ) + (\rho_\phi + p_\phi) \right] \, , \label{eq:2b} \\
    \dot{\rho}_j &=& - 3 H (\rho_j + p_j ) \, , \label{eq:2c} \\
    \ddot{\phi} &=& -3 H \dot{\phi} - m_{\phi}^2 \phi \, , \label{eq:2d}
  \end{eqnarray}
\end{subequations}
where $\kappa^2 = 8\pi G$. The dot denotes the derivative with respect to cosmic time $t$, and $H=\dot{a}/a$ is the Hubble parameter. In the equations above, the scalar field density $\rho_\phi$ and pressure $p_\phi$ are defined, respectively, as
\begin{equation}
    \rho_\phi = \frac{1}{2} \dot{\phi}^2 + \frac{1}{2} m_{\phi}^2 \phi^2 \, , \quad p_\phi = \frac{1}{2} \dot{\phi}^2 - \frac{1}{2} m_{\phi}^2 \phi^2 \, . \label{eq:polar-trans}
\end{equation}

For the SFDM part, it is convenient to use the following change of variables~\cite{Copeland:1997et,Urena-Lopez:2015odd,Urena-Lopez:2015gur,Cedeno:2017sou,Roy:2018nce,LinaresCedeno:2020dte},
\begin{subequations}
\label{eq:dynsyscart}
\begin{eqnarray}
\frac{\kappa \dot{\phi}}{\sqrt{6} H} = e^\beta \sin(\theta/2)  \, &,& \quad -\frac{\kappa m_{\phi} \phi}{\sqrt{6} H} = e^\beta \cos(\theta/2) \, , \label{eq:dynsyscart-a} \\ 
y_1 &\equiv& \frac{2m_{\phi}}{H} \, . \label{eq:dynsyscart-b}
\end{eqnarray}
\end{subequations}

To understand the meaning of the new variables, we write here the SFDM density parameter $\Omega_\phi$ and equation of state (EOS) $w_\phi$, 
\begin{subequations}
\label{eq:eos-Ophi}
\begin{eqnarray}
    \Omega_\phi &=& \frac{8\pi G \rho_\phi}{3H^2} = e^{2\beta} \, , \label{eq:eos-Ophi-a} \\
    w_\phi &=& \frac{\dot{\phi}^2 - m_{\phi}^2 \phi^2}{\dot{\phi}^2 + m_{\phi}^2 \phi^2} = -\cos \theta \, . \label{eq:eos-Ophi-b}
\end{eqnarray}
\end{subequations}

Thus, $\beta$ is the logarithm of the energy density parameter, and $\theta$, being an internal polar angle, is directly related to $w_\phi$. Lastly, $y_1$ is simply the ratio of the field mass to the Hubble parameter (i.e. dimensionless by definition), which is an ubiquitous quantity in all methods for solutions of the SFDM equations of motion. 

Using the new variables, the Klein-Gordon equation~\eqref{eq:2d} transforms into the following set of first order differential equations~\cite{Urena-Lopez:2015gur},
\begin{subequations}
\label{eq:dy_sys}
  \begin{eqnarray}
  \theta^\prime &=& -3 \sin \theta + y_1 \, , \label{eq:dy_sys-a} \\
  y^\prime_1 &=& \frac{3}{2} \left( 1 + w_{\rm tot} \right) y_1 \,
  , \label{eq:dy_sys-b} \\
  \beta^\prime &=& \frac{3}{2} (w_{\rm tot} + \cos \theta) \, . \label{eq:dy_sys-c}
\end{eqnarray}
\end{subequations}

Here, a prime denotes derivatives with respect to the number of $e$-folds of expansion $N=\ln a$. The total EOS $w_{\rm tot}$, which is used in the foregoing equations, can be calculated from the ratio of the total pressure $p_{\rm tot}$ to the total density $\rho_{\rm tot}$ in the Universe, and its explicit expression is given in terms of the EOS of the different components of matter as
\begin{equation}
    w_{\rm tot} = \frac{p_{\rm tot}}{\rho_{\rm tot}} = \frac{\sum_j w_j \rho_j + w_\phi \rho_\phi}{\sum_j \rho_j + \rho_\phi} \, .
\end{equation}

Although the new dynamical variables are convenient for numerical purposes, we need to recover the standard quantities used in Boltzmann and cosmological codes in general, for instance, the density and pressure of the SFDM component. It can be shown that such quantities can be recovered in the form
\begin{equation}
    \rho_\phi = \frac{e^{2\beta}}{1-e^{2\beta}} \sum_j \rho_j \, , \quad p_\phi = - \cos \theta \, \rho_\phi \, ,
\end{equation}
where the sum takes into account only the density components other than the SFDM one.

\subsection{\label{sec:lin-dens}Linear density perturbations}
When considering linear perturbations, the line element in the synchronous gauge, and the perturbed scalar field are given by
\begin{equation}
ds^2 = -dt^2+a^2(t)(\delta_{ij}+\bar{h}_{ij})dx^idx^j \, , \label{eq:mfpert}
\end{equation}
\begin{equation}
    \phi(\vec{x},t) = \phi(t)+\varphi(\vec{x},t) \, . \label{eq:pertphi}
\end{equation}
where $\bar{h}_{ij}$ and $\varphi$ are the spatial part of the metric perturbation and scalar field fluctuation, respectively. The linearized Klein-Gordon equation for the scalar field perturbation is written (in Fourier space) as~\cite{Ratra:1990me,Ferreira:1997au,Ferreira:1997hj,Perrotta:1998vf}
\begin{equation}
  \ddot{\varphi}(\vec{k},t) = - 3H \dot{\varphi}(\vec{k},t) - \left( m_{\phi}^2 + \frac{k^2}{a^2} \right) \varphi(\vec{k},t) - \frac{1}{2} \dot{\phi} \dot{\bar{h}} \, . \label{eq:13}
\end{equation}

In a similar way to the procedure we have done for the Klein-Gordon equation~\eqref{eq:2b}, we propose the following change of variables for the scalar field perturbation $\varphi$ and its derivative $\dot{\varphi}$ \cite{Urena-Lopez:2015gur,Cedeno:2017sou} 
\begin{equation}
\sqrt{\frac{2}{3}} \frac{\kappa \dot{\varphi}}{H} = - e^{\alpha + \beta} \cos(\vartheta/2) \, ,  \quad
    \frac{\kappa y_1 \varphi}{\sqrt{6}} = - e^{\alpha +\beta} \sin(\vartheta/2) \, . \label{eq:22a}
\end{equation}

Field perturbations are then represented by the new variables $\alpha$ and $\vartheta$. However, it is more convenient to further define two new variables in the form of
\begin{equation}
\delta_0=-e^{\alpha} \sin \left( \frac{\theta - \vartheta}{2} \right) \, , \quad \delta_1=-e^{\alpha} \cos \left( \frac{\theta - \vartheta}{2} \right) \, . \label{eq:d0d1}
\end{equation}

After some straightforward algebra, the linearized Klein-Gordon equation~\eqref{eq:13} is represented by the following set of first-order differential equations, 
\begin{subequations}
\label{eq:newdeltas}
\begin{eqnarray}
\delta^\prime_0 &=&  \left[-3\sin\theta-\frac{k^2}{k^2_J}(1 - \cos \theta) \right] \delta_1 + \frac{k^2}{k^2_J} \sin \theta \delta_0  \nonumber \\ 
&& - \frac{\bar{h}^\prime}{2}(1-\cos\theta) \, , \label{eq:newdeltas-a} \\
\delta^\prime_1 &=& \left[-3\cos \theta - \frac{k^2}{k_J^2} \sin\theta \right] \delta_1 + \frac{k^2}{k_J^2} \left(1 + \cos \theta \right) \, \delta_0 \nonumber \\
&& - \frac{\bar{h}^\prime}{2} \sin \theta \, . \label{eq:newdeltas-b}
\end{eqnarray}
\end{subequations}

In writing Eqs.~\eqref{eq:newdeltas}, there appears a natural definition of a Jeans wavenumber given by $k^2_J = a^2 H^2 y_1$, which also acts as a normalization factor of the wavenumber $k$.

Notice that for scales larger than the Jeans one, $k^2/k^2_J \ll 1$, the scale-dependent terms in Eqs.~\eqref{eq:newdeltas} can be neglected, whereas in the opposite case, $k^2/k^2_J \gg 1$, the evolution of the perturbations are different for each scale. Our definition of the Jeans wavenumber coincides with those given in the literature if we write it as $k^2_J = 2a^2 m_{\phi} H$ using $y_1 = 2m_{\phi}/H$ (see Eq.~\eqref{eq:dynsyscart-b}).

To understand the physical meaning of the new dynamical variables $\delta_0,\delta_1$, we first write their expressions in terms of the original field variables,
\begin{subequations}
    \begin{eqnarray}
        \delta_0 &=& \frac{m_{\phi}^2}{\kappa^2 \rho_\phi} \left[ \frac{\kappa \dot{\varphi}}{m_{\phi}} \frac{\kappa \dot{\phi}}{m_{\phi}} + (\kappa \varphi) (\kappa \phi) \right] \, , \\
        \delta_1 &=& \frac{m_{\phi}^2}{\kappa^2 \rho_\phi} \left[ \frac{\kappa \dot{\phi}}{m_{\phi}} (\kappa \varphi) - \frac{\kappa \dot{\varphi}}{m_{\phi}} (\kappa \phi) \right] \, .
    \end{eqnarray}
\end{subequations}

Now, we recall the expressions for the scalar field density contrast $\delta_\phi$ and the velocity divergence $\Theta_\phi$ in the form,
\begin{subequations}
\label{eq:old-to-new-deltas}
\begin{eqnarray}
    \delta_\phi &=& \frac{\dot{\phi} \dot{\varphi} + m_{\phi} \phi \varphi}{\rho_\phi} = \delta_0 \, , \label{eq:old-to-new-deltas-a} \\
    (\rho_\phi +p_\phi) \Theta_\phi &=& \frac{k^2}{a} \dot{\phi} \varphi = \frac{k^2 \rho_\phi}{a H y_1} \left[ (1 -\cos \theta) \delta_1 - \sin \theta \delta_0 \right] \nonumber \, .\\ \label{eq:old-to-new-deltas-b}
\end{eqnarray}
\end{subequations}
Hence, the variable $\delta_0$ is the scalar field density contrast, whereas the velocity divergence is based on a combination of the two variables $\delta_0$ and $\delta_1$. See Appendix~\ref{sec:fld_approx} for an extended discussion of the equivalence between our approach and the fluid approximation.

\subsection{\label{sec:numer-evol} Initial conditions}
Calculation of initial conditions implies an approximate solution of the equations of motion starting well within the epoch of radiation domination, with a corresponding initial value of the scale factor of the order of $a_i = 10^{-14}$~\cite{Urena-Lopez:2015gur}. 

We start with the initial condition of the auxiliary variable, which reads
\begin{equation}
    y_{1i} = \frac{2m_{\phi}}{H_0} \frac{H_0}{H_i} = 1.85 \times 10^{11} \frac{a^2_i}{\sqrt{\Omega_{r0} h^2}} \left( \frac{m_{\phi}}{\mathrm{10^{-22} eV}} \right) \, , \label{eq:y1initial}
\end{equation}
where $H_0$ ($H_i$) is the present (initial) value of the Hubble parameter, and $h$ is its reduced value. Note that for the calculation of $H_i$ we assume radiation domination at early times. It is clear from Eq.~\eqref{eq:y1initial} that the auxiliary variable is very small at early times, $y_{1i} \sim 10^{-14}$, for the values of $m_{\phi}$ that are of interest for SFDM models. On the contrary, its value at present is very large $y_1 \sim 10^{10}$, which means that it changes by almost 24 orders of magnitude during its evolution.

As for the polar angle, there is an atractor solution when the equations of motion are solved in the linear regime at early times, from which we obtain the following equations.
\begin{equation}
    \theta_i = \frac{1}{5} y_{1i} \, .
\end{equation}

The initial condition of the variable $\beta$ is found by matching the early and late time solutions at the beginning of the rapid oscillations of the field. The resultant equation is
\begin{equation}
    e^{2\beta_i} = A \times a_i \frac{\Omega_{\phi 0}}{\Omega_{r0}} \left[ \frac{4\theta^2_i}{\pi^2} \left( \frac{1+\pi^2/36}{1+\theta^2_i/9} \right) \right]^{3/4} \, .
\end{equation}

Here, $A$ is a constant coefficient that is adjusted by the numerical code, typically with a shooting mechanism, to match the value of the desired density parameter $\Omega_{\phi 0}$ at the present time.

On the other hand, the initial conditions of the density perturbations is a more involved procedure, but it reveals the existence of an attractor solution for the dynamical variables in the form
\begin{subequations}
\label{eq:delta_ini}
\begin{eqnarray}
    \delta_{0i} = \frac{2}{7} \bar{h} \sin(\theta_i/2) \sin(\theta_i/12) \, , \\ 
    \delta_{1i} = \frac{2}{7} \bar{h} \sin(\theta_i/2) \cos(\theta_i/12) \, .
\end{eqnarray}
\end{subequations}

The details of the numerical implementation of the polar method of the sections above in the amended version of the Boltzmann code \texttt{CLASS} are presented in Appendix~\ref{sec:class-details}, from which we obtained the numerical solutions that are presented in the sections below.

\section{The stage of rapid field oscillations \label{sec_osc}}
For the field to behave as a CDM component, it should enter a phase of rapid oscillations around the minimum of the potential. Under our polar transformation~\eqref{eq:polar-trans}, such fast oscillations are equivalent to the following averages during a Hubble time, $\langle \sin \theta \rangle = 0$ and $\langle \cos \theta \rangle = 0$. 

However, we must be careful in the form the oscillations are dealt with, as the solutions at late times depend on the choices made for the averaged dynamical quantities. Here, we explain in detail our method for the cutoff of the rapid oscillations proposed in~\cite{Urena-Lopez:2015gur,Cedeno:2017sou,LinaresCedeno:2020dte}.

\subsection{Outline of the general method}\label{sec:III-A}
It is well known that the stage of rapid oscillations is difficult to solve numerically, and then we follow here the prescription in Ref.~\cite{Urena-Lopez:2015gur} in that the cosine and sine functions in the equations of motion are replaced by the cutoff trigonometric functions.
\begin{subequations}
\label{eq:trig-cutoff}
\begin{eqnarray}
    \cos_\star \theta &=& \frac{1}{2} \left[ 1 - \tanh (\theta - \theta_\star) \right] \cos \theta \, , \\
    \sin_\star \theta &=& \frac{1}{2} \left[ 1 - \tanh (\theta - \theta_\star) \right] \sin \theta \, ,
\end{eqnarray}
\end{subequations}
where $\theta_\star$ is a reference value. In this form, $\cos_\star \theta = \cos \theta$ ($\sin_\star \theta = \sin \theta$) if $0 \leq \theta < \theta_\star$, while $\cos_\star \theta \to 0$ ($\sin_\star \theta \to 0$) if $\theta \gg \theta_\star$.

In the following, we will refer to $t_\star$ as the time at which we apply the cutoff for the trigonometric functions and then to $\theta(t_\star) = \theta_\star$. We will also refer to $t_{osc}$ as the time for the beginning of the rapid oscillations. However, and in contrast to $t_\star$, the value of $t_{osc}$ cannot be precisely determined, and in our formalism it is just a reference value without a major effect on the numerical solutions. 

The general method can be described as follows. We replace all the sine and cosine terms with the cutoff functions~\eqref{eq:trig-cutoff} in the equations of motion~\eqref{eq:dy_sys}, and then solve them numerically. Note that the solutions are then continuous at $t=t_\star$ by construction, and we only need to be sure that the cutoff is applied after the onset of the rapid oscillations so that $t_\star > t_{osc}$.

However, it is difficult to determine the time at the start of the oscillations, and this also makes impractical the calculation of $t_\star$. The reason is that cosmic time is a dimensional quantity calculated from the integration of the Friedmann equation~\eqref{eq:2a}, which depends on all dynamical variables in a cosmological model~\footnote{This is a similar problem to those in other approaches that require to set the value of the ratio $m_{\phi}/H$ at the start of the oscillations, where $H$ is also a dimensional output value of the cosmological equations.}. As we shall show in the following, it is better to calibrate the cutoff time using the polar variable $\theta$, which is dimensionless and also a direct dynamical variable in our set of field equations. 

\subsection{\label{sec:pol-ang}
The case of the polar angle $\theta$}
To understand the general behavior of the solutions after the cutoff of the trigonometric functions, we start with the equation of motion~\eqref{eq:dy_sys-a} of the polar angle $\theta$, which for convenience we write in terms of cosmic time $t$,
\begin{equation}
    \dot{\theta} = -\frac{3}{2t} \sin \theta + 2m_{\phi} \, , \label{eq:theta-equation}
\end{equation}
where we have considered $H = 1/(2t)$ for RD. Notice that we can write Eq.~\eqref{eq:theta-equation} in the form
\begin{equation}
    \frac{d\theta}{d(2m_{\phi} t)} = -\frac{3}{2} \frac{\sin \theta}{(2m_{\phi}t)} + 1 \, , \label{eq:theta-dimensionless}
\end{equation}
which shows that the evolution of $\theta$ as a function of the dimensionless variable $2m_{\phi} t$ is the same, regardless of the value of the field mass $m_{\phi}$. We will use this feature in the plots below, but we still refer to Eq.~\eqref{eq:theta-equation} to obtain semi-analytical expressions.

First, we assume at the beginning that $0 < \theta \ll 1$, and then $\sin \theta \simeq \theta$. As a consequence, Eq.~\eqref{eq:theta-equation} becomes
\begin{subequations}
\label{eq:theta-first}
\begin{equation}
    \dot{\theta} = - \frac{3}{2t} \theta + 2m_{\phi} \, , \label{eq:theta-first-a}
\end{equation}
and the solution that satisfies the initial condition $\theta(0)=0$ is 
\begin{equation}
    \theta(t) = \frac{4}{5} m_{\phi} t \, . \quad \mathrm{(Early)} \label{eq:theta-first-b}
\end{equation}
\end{subequations}

We now use Eq.~\eqref{eq:theta-first-b} on the right-hand side of Eq.~\eqref{eq:theta-equation}, from which we obtain the new differential equation,
\begin{subequations}
\label{eq:theta-second}
\begin{equation}
    \dot{\theta} = - \frac{3}{2t} \sin_\star(4m_{\phi} t/5) + 2m_{\phi} \, , \label{eq:theta-second-a}
\end{equation}
whose solution is
\begin{equation}
    \theta (t) = 2m_{\phi} t - \frac{3}{2} \mathrm{Si}(4m_{\phi} t/5) \, , \label{eq:theta-second-b}
\end{equation}
where $\mathrm{Si}(x)$ is the sine integral. It can be shown that, at early times $m_{\phi} t \ll 1$, we recover the solution~\eqref{eq:theta-first-b}. Likewise, for the late-time evolution we can approximate the sine integral by its asymptotic behavior, $\mathrm{Si}(x) \simeq \pi/2 - \cos(x)/x + \mathcal{O}(1/x^2)$ for $x \gg 1$, to obtain
\begin{equation}
    \theta (t) = 2m_{\phi} t - \frac{3\pi}{4} \, . \quad \mathrm{(Late)} \label{eq:theta-late-ba}
\end{equation}
Any further iteration to integrate Eq.~\eqref{eq:theta-equation} cannot be expressed in closed form, but, as we present below, Eq.~\eqref{eq:theta-second-b} suffices to analyze the main properties in the time evolution of the polar angle $\theta$.
\end{subequations}

\begin{figure}[htp!]
\includegraphics[width=0.49\textwidth]{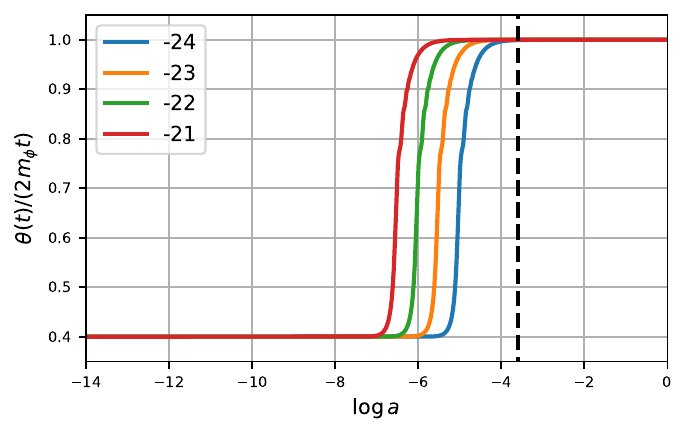}
\includegraphics[width=0.49\textwidth]{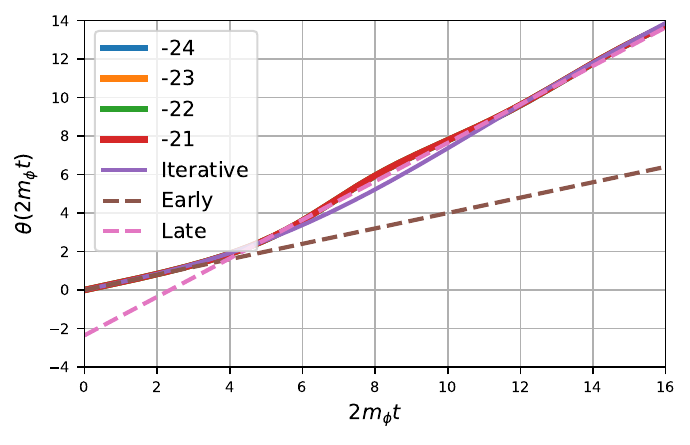}
\caption{\label{fig:polar1} Numerical solutions of the polar variable $\theta$ as a function of the dimensionless variable $2m_{\phi} t$. The cases correspond to different values of the field mass $\log(m_{\phi}/\mathrm{eV}) = -24, -23,-22,-21$. Also shown are the semi-analytical solutions~\eqref{eq:theta-second-b} (Iterative, purple),~\eqref{eq:theta-first-b} (Early, dashed brown), and~\eqref{eq:theta-late-ba} (late, dashed pink). See the text for more details.}
\end{figure}

The numerical solutions of Eq.~\eqref{eq:theta-equation} for different values of the field mass $m_{\phi}$, and as a function of the scale factor $a$, are shown in the top panel of Fig.~\ref{fig:polar1}. It can be seen that the polar variable shows two asymptotic behaviors that correspond to the semi-analytical solutions: at early times $\theta/(2m_{\phi} t) \to 5/4$, while at late times $\theta/(2m_{\phi} t) \to 1$, as indicated by Eqs.~\eqref{eq:theta-first-b} and~\eqref{eq:theta-late-ba}, respectively. These asymptotic limits are the same for any field mass $m_{\phi}$, and the latter only influences the time at which the transition occurs between the two values.

In the bottom panel of Fig.~\ref{fig:polar1} we have the evolution of the polar variable but now in terms of dimensionless cosmic time $2m_{\phi} t$. Although the numerical solutions are shown in different colors, it is clear that the corresponding curves are superimposed on each other because their behavior is the same.

It can be seen that the semi-analytical solutions agree well with the numerical ones. In particular, the iterative solution~\eqref{eq:theta-second-b} gives a reliable description of the early and late time trends of the solutions, and it even gives a good approximation to the oscillations of the numerical solutions at intermediate times $2m_{\phi} t \simeq 4$, which is also the time at which $\theta \simeq \pi/2$. That is, it also corresponds to the time at which the scalar field EOS first crosses the zero value $w_\phi \simeq 0$. As this occurs within radiation domination, we also find $2m_{\phi} t = m_{\phi}/H \simeq 4$, which is the typical time for the start of the oscillations estimated for these field systems.

Surprisingly, the bottom panel of Fig.~\ref{fig:polar1} also shows that the late-time expression~\eqref{eq:theta-late-ba} also seems to work very well from the intermediate times onward, that is, almost from the start of the rapid oscillations. This means that we can safely write
\begin{subequations}
\label{eq:theta-star}
\begin{equation}
    \theta(t > t_{osc}) = 2m_{\phi} t - \frac{3\pi}{4} \, . \label{eq:theta-star-a}
\end{equation}

That Eq.~\eqref{eq:theta-star-a} is also a very good approximation can be understood from the properties of the sine integral $\mathrm{Si}(x)$, which rapidly converges to its asymptotic value of $\pi/2$, with small oscillations around it that rapidly decay away. In what follows, we will use Eq.~\eqref{eq:theta-star-a} to describe the behavior of the polar angle after the onset of rapid oscillations of the field $\phi$.

We can also use Eq.~\eqref{eq:theta-star-a} also to convert the cutoff time $t_\star$ into a cutoff polar angle $\theta_\star$, which is both a dynamical variable and the argument in the modified trigonometric functions~\eqref{eq:trig-cutoff}. Hence, the relation between the cutoff values $t_\star$ and $\theta_\star$ is
\begin{equation}
    2m_{\phi} t_\star = \theta_\star + \frac{3\pi}{4} \, . \label{eq:theta-star-b}
\end{equation}
\end{subequations}

Equations~\eqref{eq:theta-star} are a central result in the description of our method. First, Eq.~\eqref{eq:theta-star-a} shows that the polar angle evolves linearly with cosmic time $t$ after the cutoff time. Second, Eq.~\eqref{eq:theta-star-b} allows us to determine the cutoff point of rapid oscillations via the polar angle $\theta_\star$, which is more convenient from the numerical point of view and justifies the use of the cutoff expressions~\eqref{eq:trig-cutoff}.

\begin{figure}[htp!]
\includegraphics[width=0.49\textwidth]{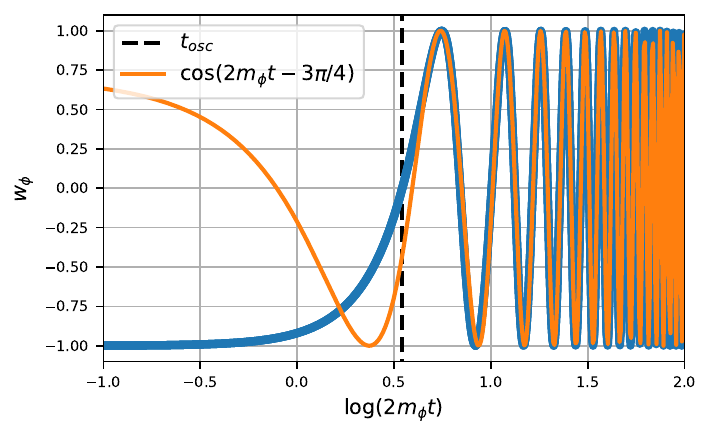}
\caption{\label{fig:EOS}The numerical solution of the scalar field EOS $w_\phi$ as a function of the variable $2m_{\phi} t$, under which all the different cases collapse into a single curve (blue curve). The dashed vertical lines mark the beginning of the rapid field oscillations at $t_{osc}$. The curve of the semi-analytical approximation~\eqref{eq:wphi-osc} (orange curve), which is well matched to the numerical solutions at $t > t_{osc}$, is also plotted. See the text for more details.}
\end{figure}

To finish this section, in Fig.~\ref{fig:EOS} we show the numerical evolution of the scalar field EOS $w_\phi$ as a function again of the dimensionless variable $2m_{\phi} t$, and we see that it first passes through zero (for $\theta = \pi/2$) at around the time $2m_{\phi} t_{osc} \simeq 3.47$, which we use to mark the time $t_{osc}$ for the start of the rapid oscillations. Notice that in terms of the usual mass-to-Hubble ratio, this is equivalent to $m_{\phi}/H_{osc} \simeq 3.47$, a value used as a reference in other studies of field models.

For comparison, we also plot in Fig.~\ref{fig:EOS} the result of the expression $\cos(2m_{\phi} t -3\pi/4)$. Notice that there is very good agreement of this curve with the original EOS $w_\phi$ almost from the start of the rapid oscillations, which means that we can use the following expression for the field EOS,
\begin{equation}
    w_\phi (t > t_{osc}) = - \cos \left( 2m_{\phi} t - 3\pi/4 \right) \, . \label{eq:wphi-osc}
\end{equation}
Equation~\eqref{eq:wphi-osc} agrees with the common wisdom that the EOS oscillates with a frequency directly related to the field mass via $2m_{\phi}$. The phase of $3\pi/4$ becomes negligible at very late times, but as we shall see, it must be taken into account for a correct description of the dynamics at intermediate times of other variables after the onset of the rapid oscillations.

\subsection{The case of the energy density $\rho_\phi$}
We have found semi-analytical results to follow the evolution of the polar variable $\theta(t)$, which are in good agreement with the numerical results. However, one should worry about the numerical accuracy as the scalar field equations of motion must be solved together with other matter components in Boltzmann codes, covering an ample time interval for a complete description of diverse cosmological phenomena.

Here, we perform some accuracy tests using the amended version of the Boltzmann code \texttt{CLASS}, taking some guidelines from our semi-analytical results. Our main concern is the choice of the cutoff value $\theta_\star$. As we shall see, the cutoff procedure leaves a residual difference with respect to the expected late-time evolution of a given variable that can be minimized if $\theta_\star \gg 1$.

An example of the effects of the cutoff on the evolution of different quantities is the energy density $\rho_\phi$, which obeys the equation
\label{eq:density}
\begin{equation}
    \dot{\rho}_\phi = -3 H (1+w_\phi) \rho_\phi = -3 H \left( 1 - \cos \theta \right) \rho_\phi \, , \label{eq:density-a}
\end{equation}
whose formal solution after the onset of the rapid oscillations can be written as
\begin{subequations}
\begin{equation}
    \rho_\phi a^3 = \rho_{\phi, osc} a^3_{osc} \exp \left[ F(t)-F(t_{osc}) \right] \, , \label{eq:density-b}
\end{equation}
with
\begin{equation}
   F(t) - F(t_{osc}) = 3 \int^t_{t_{osc}} H(x) \cos (\theta (x)) \, dx \, . \label{eq:density-c}
\end{equation}
\end{subequations}

It suffices to understand the behavior of the density before the time of radiation-matter equality, and for that we proceed as follows. Equation~\eqref{eq:density-c} can be written in a more convenient form if we use Eq.~\eqref{eq:theta-star-a} for the evolution of the polar angle, and then it can be shown that
\begin{subequations}
\begin{equation}
    F(t) = \frac{3\sqrt{2}}{4} \left[ \mathrm{Si} (2m_{\phi}t) - \mathrm{Ci} (2m_{\phi}t) \right] \, , \label{eq:Ffunction}
\end{equation}
where $\mathrm{Si}(x)$ and $\mathrm{Ci}(x)$ are the sine and cosine integrals, respectively. 

We are interested in the evolution of the density at late times, that is, $2m_{\phi} t \gg 1$. Given the asymptotic properties of the sine and cosine integrals for $x \gg 1$, $\mathrm{Si}(x) \simeq \pi/2 - \cos(x)/x$ and $\mathrm{Ci}(x) \simeq \sin(x)/x$, and substituting the polar angle $\theta$ using Eq.~\eqref{eq:theta-star-a}, we find that
\begin{equation}
    F(t \gg t_{osc}) \simeq \frac{3 \sqrt{2} \pi}{8} + \frac{3\sin \theta}{2(\theta + 3\pi/4)} \, . \label{eq:Ffunction-a}
\end{equation}
\end{subequations}

The last term in Eq.~\eqref{eq:Ffunction-a} will be responsible for a residual oscillatory term in the density, which will decay away. In fact, if we define
\begin{equation}
    \rho_{\phi 0} \equiv \rho_{\phi, osc} a^3_{osc} \exp \left[ \frac{3 \sqrt{2} \pi}{8} -F(t_{osc}) \right] \, , \label{eq:rho0}
\end{equation}
we can also write Eq.~\eqref{eq:density-b} in a more neat form as
\begin{equation}
    \rho_\phi (t \gg t_{osc}) = (\rho_{\phi 0} /a^3) \exp \left( \frac{3\sin \theta}{2(\theta + 3\pi/4)} \right) \, , \label{eq:rho-after-osc}
\end{equation}
which is correct for $\theta \gg 1$. There are two parts in the rhs of Eq.~\eqref{eq:rho-after-osc}: one that evolves steadily at the same rate as a pressureless component $(\sim a^{-3})$, and another one that contributes with a decaying oscillating term around unity provided by the exponential function. Moreover, $\rho_{\phi 0}$ represents the correct asymptotic value of the field density at late times.

The following question arises: can we be assured that our cutoff procedure of the rapid oscillations recovers the right evolution of the density at late times? First, notice that, in principle, $\rho_{\phi 0}$ in Eq.~\eqref{eq:rho0} is fixed at the onset of the rapid oscillations, but we do not need to be very specific about the values of the parameters at this time. In our method, unlike others in the literature, we do not require knowing the precise value of $t_{osc}$, and we can be completely oblivious to it as long as we ensure $t_\star > t_{osc}$. The reason is simple: the cutoff of the rapid oscillations is made smoothly at the level of the equations of motion, and then there is no loss of continuity in the numerical variables.

To answer our question above, Eq.~\eqref{eq:rho-after-osc} should be compared with the truncated case. After the cutoff, the equation of motion for the density is
\begin{subequations}
\label{eq:rho-after}
\begin{equation}
    \dot{\rho}_\phi = -3 H \rho_\phi \, , \label{eq:rho-after-a}
\end{equation}
whose solution simply is $\rho (t > t_\star) = \rho_\star/a^3$, where $\rho_\star$ is the density value at $t=t_\star$. By the continuity of the solutions at $t_\star$ for Eqs.~\eqref{eq:density-a} and~\eqref{eq:rho-after-a}, we finally get
\begin{equation}
    \rho (t > t_\star) = (\rho_{\phi 0} /a^3) \exp \left( \frac{3\sin \theta_\star}{2(\theta_\star + 3\pi/4)} \right) \, . \label{eq:rho-after-b}
\end{equation}
\end{subequations}
The result is quite direct: the cutoff introduces a small mismatch, and the correct asymptotic value of the density is not recovered from the solution~\eqref{eq:rho-after-b}. But the discrepancy depends on the cutoff value $\theta_\star$, and, in principle, it can be made as small as required if $\theta_\star \gg 1$. 

However, Eq.~\eqref{eq:rho-after-b} itself suggests a faster route, which is to choose $\theta_\star = n\pi$, where $n$ is an integer number, although a large enough one so that still $\theta_\star \gg 1$, as this also allows us to neglect other oscillatory terms in the sine and cosine integrals in Eq.~\eqref{eq:Ffunction} that are of order $\mathcal{O}(1/\theta^2_\star)$ and smaller.

\begin{figure*}[htp!]
\includegraphics[width=0.49\textwidth]{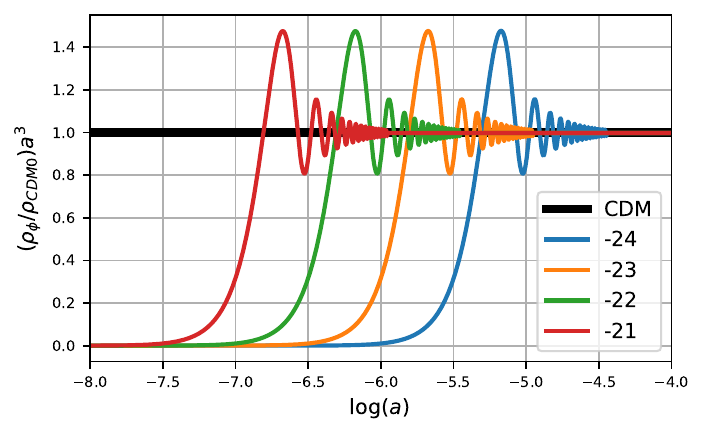}
\includegraphics[width=0.49\textwidth]{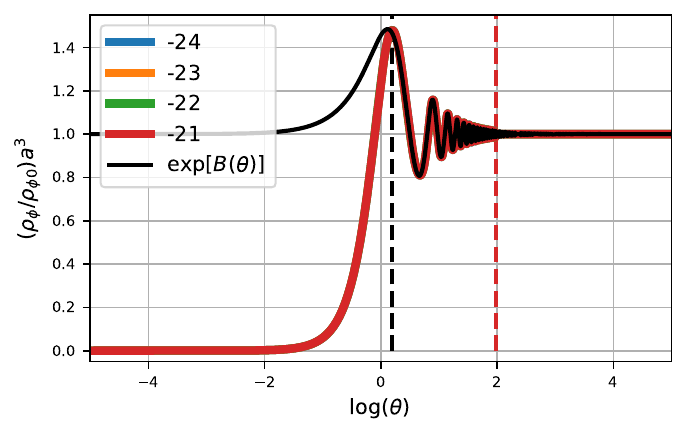}
\caption{\label{fig:rho}Numerical solutions of the scalar field energy density $\rho_\phi$, in a combination that highlights the asymptotic, non-oscillatory value at late times; the curves represent the same cases as in Fig.~\ref{fig:polar1}. (Left) The energy density, as a function of the scale factor $a$, behaves as a pressureless component from the onset of rapid oscillations, which occur at times that depend on the field mass $m_{\phi}$, but the late-time behavior is the same: decaying oscillations around a fixed value. The cutoff value was set at $\theta_\star = 30\pi$ for the four cases. (Right) If plotted as a function of the polar angle $\theta$, all curves collapse into a single one. Also shown is the semi-analytical formula~\eqref{eq:rho-after-osc}, with $B(\theta) \equiv 3\sin \theta/(\theta + 3\pi/4)$ (black curve), which is in good agreement with the numerical solutions from the start of the rapid oscillations at $\theta_{osc} = \pi/2$ (black dashed line). The cutoff value $\theta_\star = 30\pi$ (red dashed line) is also shown as a reference. See the text for more details.}
\end{figure*}

Numerical examples of the evolution of the field density, in the combination $(\rho_\phi/\rho_{\phi 0}) a^3$, are shown in Fig.~\ref{fig:rho}, for different values of the field mass $m_{\phi}$ but with a fixed value $\theta_\star = 30\pi$. Here, $\rho_0$ is the last value in each of the numerical solutions. Note that the asymptotic value is always unity. In the left panel, we see that the density makes a transition to a pressureless behavior once the rapid oscillations start, but as before, the transition time depends on the field mass $m_{\phi}$. 

If the density is plotted as a function of the variable $\theta$, as in the right panel of Fig.~\ref{fig:rho}, we find that all curves collapse again into a single one, and there is a common evolution for all cases. Moreover, we also show the curve arising from Eq.~\eqref{eq:rho-after-osc} (denoted by $\exp [B(\theta)]$), and it can be seen that it quite well matches the numerical curves after the onset of the rapid oscillations.

\begin{figure*}[htp!]
\includegraphics[width=0.49\textwidth]{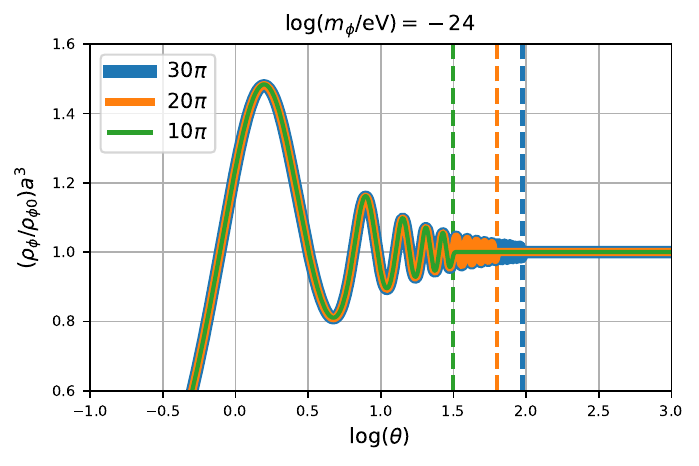}
\includegraphics[width=0.49\textwidth]{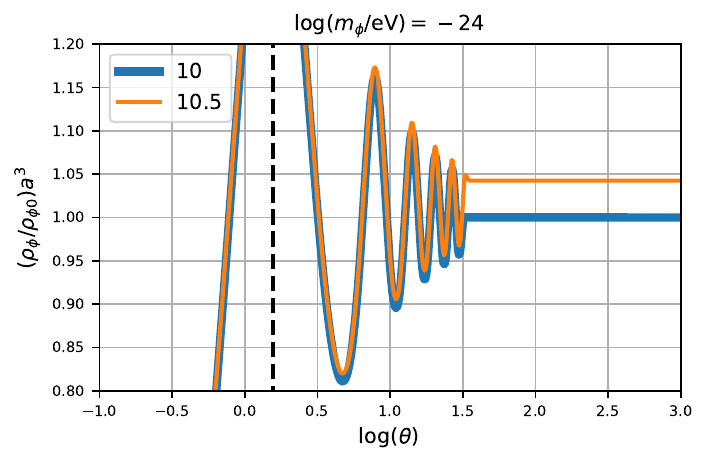}
\caption{\label{fig:rhon}The behavior of the normalized background density for different choices of the cutoff angle $\theta_\star$. The density at late times matches the average value well if $\theta_\star$ is an integer multiple of $\pi$ (left panel), while in any other cases there is a noticeable mismatch with respect to the average value (right panel). The dashed vertical lines mark the corresponding cutoff values $\theta_\ast$ of the numerical solutions. The field mass was fixed at $m_{\phi} = 10^{-24} \, \mathrm{eV}$ in the numerical examples. See the text for more details.}
\end{figure*}

Now, in Fig.~\ref{fig:rhon} we show the effects arising from different choices of the cutoff value $\theta_\star$ and with a fixed mass $m_{\phi} = 10^{-24} \mathrm{eV}$. In the left panel, we take $\theta_\star =10\pi,20\pi,30\pi$, and we see that the late-time evolution is the same for all cases (the curves are superimposed on each other) even though the cutoff of the oscillations appears at different times. Also, the correct asymptotic value of the density is recovered, and in all cases it corresponds to the expected average of the density oscillations.

For the right panel in Fig.~\ref{fig:rhon} we take the cutoff values $\theta_\star = 10\pi, 10.5\pi$. To have a good matching of the first density oscillations in the two cases and to facilitate the comparison of their asymptotic values, we applied the correction (exponential) factor that appears in Eq.~\eqref{eq:rho-after-osc} to the case $\theta_\star = 10.5\pi$.~\footnote{It must be noticed that the value of the present density $\rho_0$ is the right one in all the numerical cases, as \texttt{CLASS} adjusts accordingly the initial conditions of the dynamical variables to make the numerical solution recover the same final result. This is the reason why we needed to match the early oscillations for a fair comparison at late times between the two cases $\theta_\star = 10\pi, 10.5\pi$.} It can be seen that the cutoff occurs at the maximum of the last oscillation, and hence the asymptotic value is larger than the correct one. 

We can give an estimate of the error between the two asymptotic values, which according to Eq.~\eqref{eq:rho-after-b} is
\begin{subequations}
\begin{equation}
    \frac{\Delta \rho_{\phi 0}}{\rho_{\phi 0}} = \exp \left( \frac{3\sin \theta_\star}{2(\theta_\star + 3\pi/4)} \right) -1 \, . \label{eq:rho0-err}
\end{equation}
For the particular case with $\theta_\star = 10.5\pi$ we get
\begin{equation}
  100 \times \left[ \exp \left( \frac{3\sin (10.5\pi)}{22.5\pi} \right) -1 \right] = 4.3\% \, .
\end{equation}
\end{subequations}
This difference is not negligible if one desires high precision of the solution, and it clearly illustrates the necessity to choose the cutoff value $\theta_\star$ wisely.

\subsection{Linear density perturbations}
In contrast to the background quantities in the sections above, the analysis of Eqs.~\eqref{eq:newdeltas} is much more involved because the evolution of the quantities $\delta_0$ and $\delta_1$ is coupled to that of the so-called metric continuity $\bar{h}^\prime/2$ through the perturbed Einstein equations. It is not possible to make a clear separation of the oscillatory and non-oscillatory terms in the formal solution, and a wise decision on the cutoff value $\theta_\star$ cannot easily be decided.

\begin{figure*}[htp!]
\includegraphics[width=0.49\textwidth]{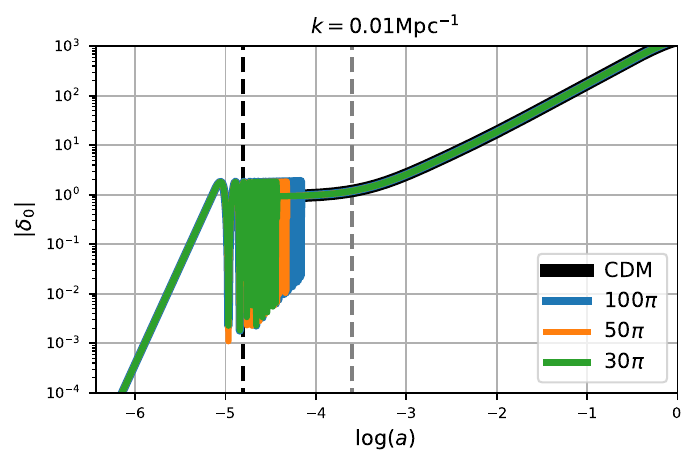}
\includegraphics[width=0.49\textwidth]{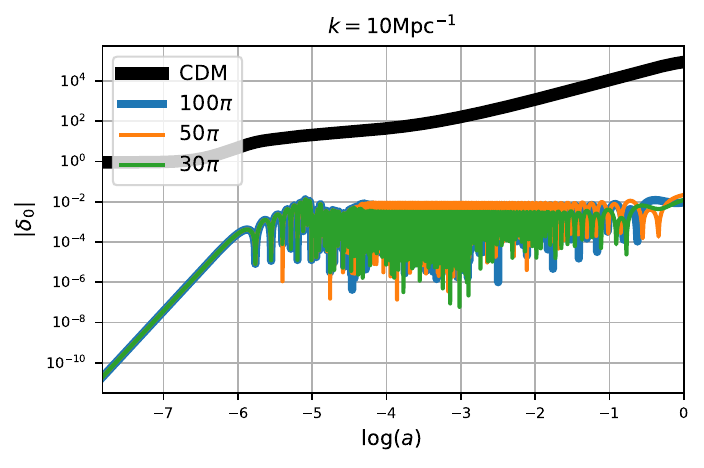}
\includegraphics[width=0.49\textwidth]{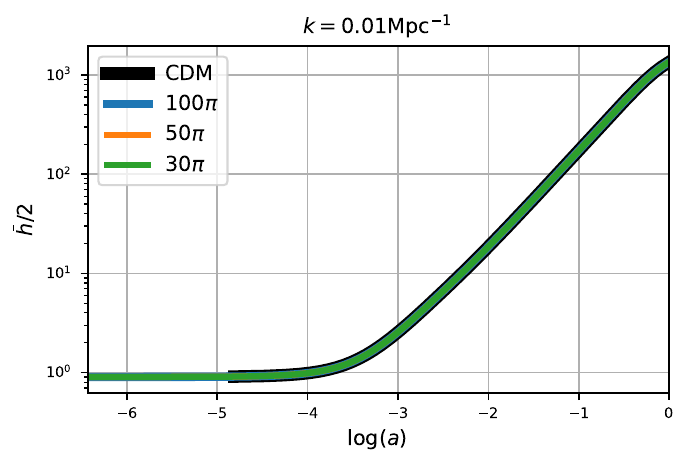}
\includegraphics[width=0.49\textwidth]{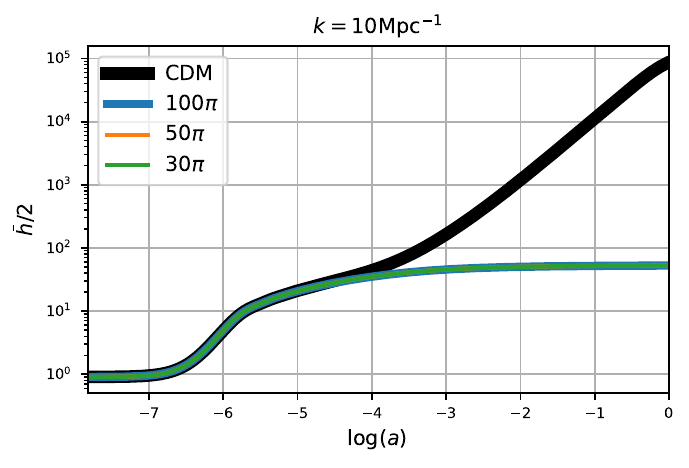}
\caption{\label{fig:deltas}Behavior of the field density contrast $\delta_0$ (top panels) and of the metric continuity $\bar{h}/2$ (bottom panels), for the different wavenumbers $k$ in the plots and for a selection of values for $\theta_\star$. It is clear that the behavior of both quantities is the same as that of their counterparts in the CDM for large scales (here represented by $k=0.01/\mathrm{Mpc}$), whereas for small scales ($k=10/\mathrm{Mpc}$) they are considerably smaller than those of CDM at late times. The field mass was fixed at $m_{\phi} = 10^{-24} \, \mathrm{eV}$ in the numerical examples. See the text for more details.}
\end{figure*}

In Fig.~\ref{fig:deltas} we show the evolution of the density contrast $\delta_0$ for a scale much larger than the Jeans wavenumber, so that $k^2 \ll k^2_J$. In the upper panels, we see the cases with the same cutoff values $\theta_\star$ used previously in Fig.~\ref{fig:rhon}. We can see that the numerical solutions have a larger stage of rapid oscillations for larger values of $\theta_\star$, as in the background case. Also, the choice $\theta_\star = n\pi$, see Eq.~\eqref{eq:rho-after-b} and the text below, does not make the numerical solution coincide with the nonoscillatory solutions at $t_\star$. There is a small, but visible, mismatch between the solutions for the different values of $\theta_\star$ considered in the graphs.

Nevertheless, the numerical solution is able to catch up with the non-oscillatory solution after the cutoff of the rapid oscillations, and this is because of the structure of the system~\eqref{eq:newdeltas}: its solution is driven by the nonhomogeneous term involving the metric continuity $\bar{h}^\prime$, which acts as an attractor solution even at early times. At late times, the value of $|\delta_0|$ oscillates around $\bar{h}/2$ with an amplitude that does not decay. It is only on average that the scalar field density contrast can be identified with the CDM one, $\langle \delta_0 \rangle = \delta_{CDM}$.

Other intrinsic oscillations, which we refer to as scale oscillations, are noticeable for scales smaller than the Jeans scale: $k^2/k^2_J \gtrsim 1$, which appear even after the cutoff of the rapid oscillations. This is because the term $k^2/k^2_J$ plays the role of a frequency in terms of the number of $e$-folds $N$ in Eqs.~\eqref{eq:newdeltas} and not in cosmic time $t$. For such small scales, from the very beginning there may be a combination of rapid oscillations with scale oscillations, and the choices of $\theta_\star$ at the transition time give different results for $\delta_0(t_\star)$ and $\delta_1(t_\star)$. 

Our numerical results for the density contrast $\delta_0$ for small scales are shown in the right panels of Fig.~\ref{fig:deltas}, where we see noticeable differences in the late-time behavior of the solutions. It is clear that it is necessary to follow the numerical solutions for longer before cutting off the rapid oscillations and to achieve some convergence of the solutions. However, the evolution of metric continuity $\bar{h}/2$ is always smooth and the same regardless of the cutoff value $\theta_\star$, although its amplitude is also highly suppressed with respect to the CDM case.

\begin{figure}[htp!]
\includegraphics[width=0.49\textwidth]{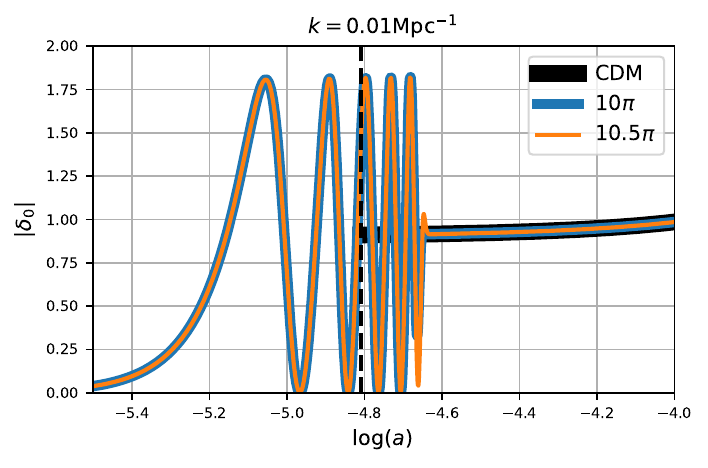}
\caption{\label{fig:deltas1}Illustration of the behavior of the density contrast $\delta_0$, for one large scale and for different choices of the cutoff angle $\theta_\star$. Although there is a mismatch between the solutions at the cutoff time, both follow exactly the CDM solution at late times. The dashed vertical line marks $\log(a_{eq})$. The field mass was fixed to $m_{\phi} = 10^{-24} \, \mathrm{eV}$ in the numerical examples. See the text for more details.}
\end{figure}

We also present the cases $\theta_\star = 10\pi,10.5\pi$ in Fig.~\ref{fig:deltas1}. Both cases show that at the cutoff time none of the numerical solutions agree with the CDM solution, but they join it quickly because of the driving term $\bar{h}^\prime/2$ in Eqs.~\eqref{eq:d0d1}. 

In summary, the numerical solutions clearly show that the scalar field density contrast behaves on average like the CDM one on large scales, and that the choice $\theta_\star = n\pi$ only helps a little for the numerical solution to have a smooth transition at the cutoff time $t_\star$. There are no further consequences, because the attractor character of the equations of motion for the density perturbations eventually leads to the right numerical result~\footnote{See also Appendix~\ref{sec:fluid-perts} for the equivalent fluid equations of motion of the density perturbations after the cutoff of the rapid oscillations. There the fluid prescription shows more clearly that SFDM density perturbations grow similarly to CDM under the condition $k < k_J$.}.

\subsection{Mass power spectrum and temperature anisotropies}
\begin{figure*}[htp!]
\includegraphics[width=0.9\textwidth]{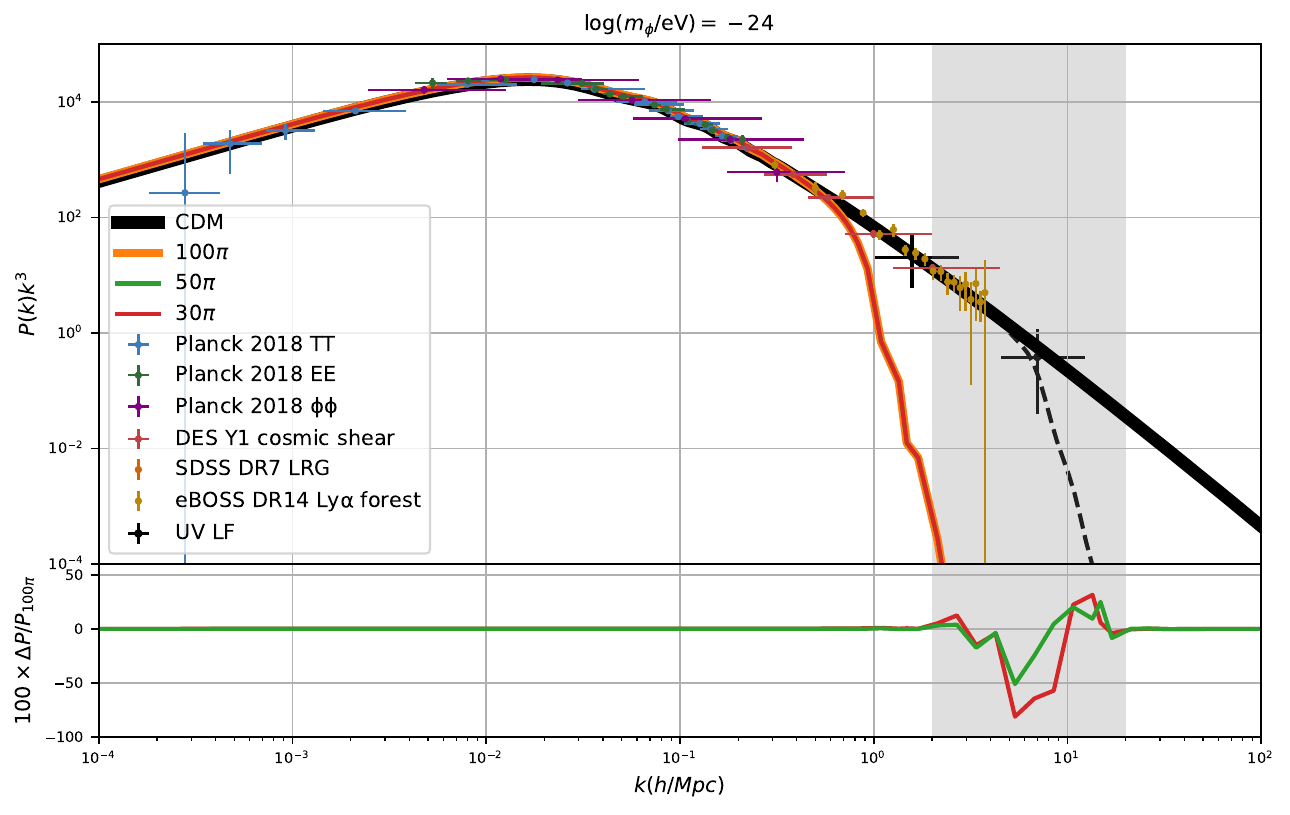}
\includegraphics[width=0.9\textwidth]{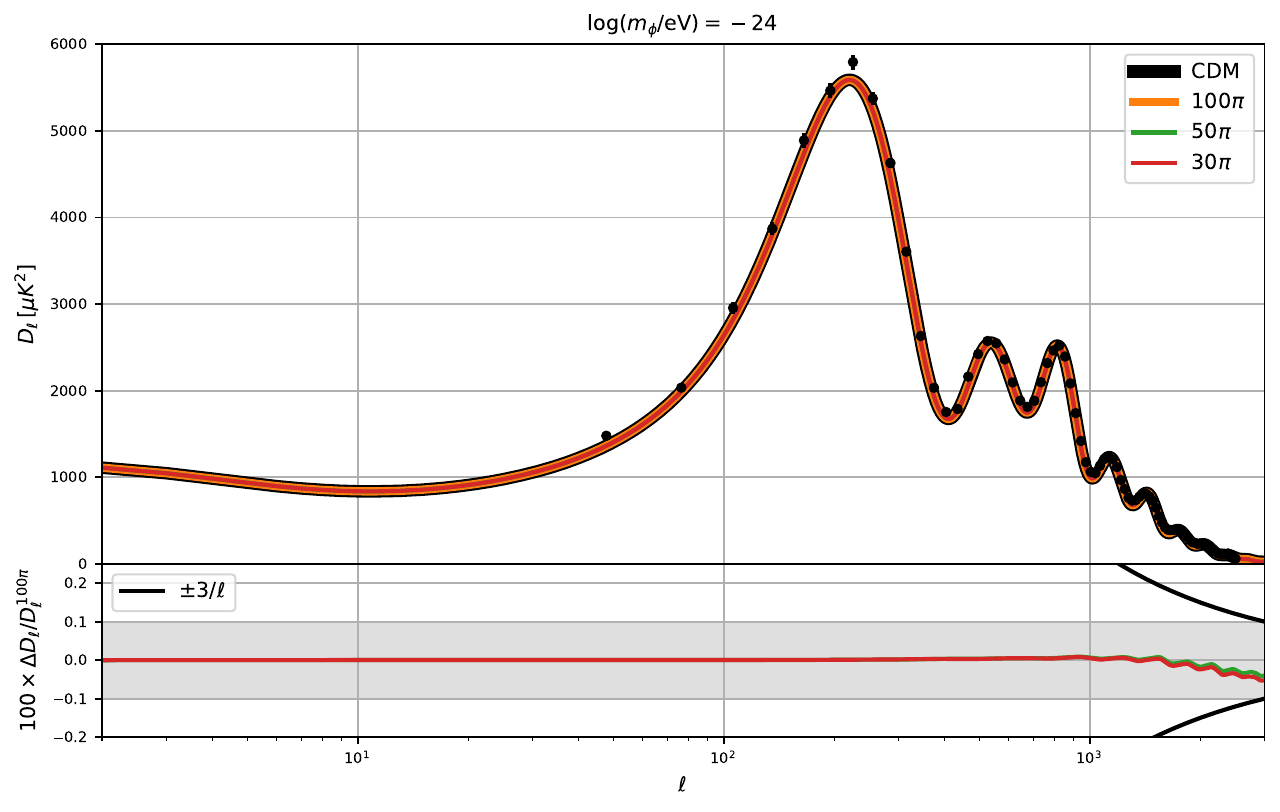}
\caption{\label{fig:deltas2} MPS $P(k)$ (top panel) and temperature spectrum $D_\ell$ (bottom panel) calculated for the field mass of $m_{\phi} = 10^{-24} \, \mathrm{eV}$, together with their relative differences with respect to MPS $P_{100\pi}(k)$ and the CDM temperature spectrum $D^{100\pi}_\ell$. The labels of the different curves refer to the cutoff variable $\theta_\star$. Note that the difference for MPS given the choices of $\theta_\star$ can be as large as $100\%$ in the range $k=2-20 \, h/\mathrm{Mpc}$ (vertical gray-shaded region), while for the temperature spectrum it is less than $0.1\%$ (horizontal gray-shaded region). For reference, we include in the temperature spectrum the (black) curves given by $\pm 3/\ell$, which represents an estimated precision threshold beyond which the parameter biases may be significant. The field mass was fixed to $m_{\phi} = 10^{-24} \, \mathrm{eV}$ in the numerical examples, but the black dashed line is for $m_{\phi} = 10^{-22} \, \mathrm{eV}$. The data points for the MPS are the measurements from the Planck 2018 CMB~\cite{Planck:2018vyg}, DES cosmic shear~\cite{DES:2017qwj}, SDSS galaxy clustering~\cite{Reid_2010}, SDSS Lyman-$\alpha$~\cite{SDSS:2017bih}, and UV LF~\cite{Sabti_2022a} data sets. See the text for more details.}
\end{figure*}

To study the issue of convergence in the solution of density perturbations for all scales, we plot the resultant mass power spectrum (MPS) in the upper panel of Fig.~\ref{fig:deltas2} the resultant mass power spectrum (MPS) for different choices of $\theta_\star$, and also the relative differences in the numerical solutions. The field mass in these examples was fixed at $m_{\phi} = 10^{-24} \, \mathrm{eV}$. The first thing to notice is that there is complete agreement with the MPS of CDM for large scales, represented by the wavenumbers $k < 0.5 \, h /\mathrm{Mpc}$.

Regarding the convergence of numerical solutions for different choices of the cutoff value $\theta_\star$, it can be seen that there is complete agreement for almost all scales, except for the interval $k=2-20 \, h/\mathrm{Mpc}$, where the difference can be as large as $100\%$. However, these discrepancies appear once the MPS is greatly suppressed compared to the CDM, although it should be noticed that the agreement is recovered once the MPS reaches a steady stage at small scales of the form $P(k) \sim k^{-3}$. Although one could use the solution with the highest cutoff value $\theta_\star$, the overall conclusion is that one can safely take $\theta_\star = 30\pi$ for reliable results with the additional advantage of saving computational time.

We repeat the comparison between different cutoff values $\theta_\star$ for the case of the temperature spectrum in terms of the variable $D_\ell = \ell(\ell +1) C_\ell/2\pi$ in the lower panel of Fig.~\ref{fig:deltas2}. This time the solutions are more similar, among themselves with different resolutions and also with the corresponding spectrum of CDM. This can be verified in the lower graph with the relative differences of the solutions with respect to choice $\theta_\star =100\pi$, which are very small and below $0.1\%$. For reference, we also show the estimated rule of thumb for bias-free parameter inference, which is given by the curves $\pm 3/\ell$ at large $\ell$. We can see that our numerical results are consistent with such a constraint for the cutoff values chosen for $\theta_\star$ (see also~\cite{PhysRevD.101.023501,PhysRevD.68.083507,MeirShimon_2013,MeirShimon_2012}).

\subsection{Constraints on $m_\phi$ from the matter power spectrum}
Here, we describe possible constraints of SFDM models from the matter power spectrum, according to recent estimates of the UV galaxy luminous function~\cite{Sabti_2022,Sabti_2022a} with the package \texttt{Gallumi}~\footnote{\texttt{https://github.com/NNSSA/GALLUMI\_public}}, and the effective field theory of the large scale structure (EFTofLSST) as in~\cite{Simon:2022csv,DAmico:2020kxu} with the package \texttt{PyBird}~\footnote{\texttt{https://github.com/pierrexyz/pybird}. The code \texttt{PyBird} has been calibrated to work on CDM models, which do not predict a cutoff in the MPS. We then assume that its EFTofLSS formalism can be extended to SFDM models in a first approximation, a possibility that seems to work well according to other similar studies~\cite{Lague:2021frh,Li_PhysRevD.99.063509} (see also~\cite{Tsedrik:2022cri,Semenaite:2022unt,Glanville:2022xes} for other beyond $\Lambda$CDM examples).}, respectively. These two are likelihoods of a recent addition to the MCMC software \texttt{MontePython}~\cite{Brinckmann:2018cvx,Audren:2012wb}, which are capable of exploring the power spectrum at semilinear scales and can be as competitive as those of Lyman-$\alpha$ observations. 

We replicated the studies in~\cite{Sabti_2022a,Simon:2022csv,Simon:2022csv} with some shortcuts, as our aim was to focus our attention on the constraints on the field mass $m_\phi$. For the analysis, we used Gaussian priors on the following parameters: the angular scale for the sound horizon with $100\theta_s =1.0411$ and $\sigma_{\theta_s} = 0.0003$~\footnote{The original study in~\cite{Sabti_2022a} fixed the value of $100\theta_s=1.0411$, as the shooting method that code \texttt{CLASS} uses to calculate it by iterations of the Hubble parameter is very efficient for the $\Lambda$CDM model. However, the same method usually does not converge in other alternative models. We decided to use a Gaussian prior to avoid failed attempts of the \texttt{CLASS} code under the command of \texttt{MontePython}.}, and the physical density of baryons $\omega_\mathrm{b} = 0.02233$ with $\sigma_\mathrm{b} = 0.00036$~\footnote{This is the same Gaussian prior of~\cite{Sabti_2022a} on the physical density of baryons, which is obtained from measurements of the abundance of primordial deuterium.}. Fixing the sound horizon $\theta_s$ is known to also fix the combination $\Omega_m h^{3.4}$, with $\Omega_m$ the physical density of total matter and $h$ the reduced Hubble constant~\cite{Sabti_2022a,2dFGRSTeam:2002tzq,Kable_2019}. This means that our Gaussian prior on $100\theta_s$, with $h=0.657$, acts as an indirect prior on the combination $\Omega_m = \Omega_b + \Omega_\phi$. For the field mass $m_\phi$, we considered a flat prior on the logarithmic scale in the range $\log(m_\phi/\mathrm{eV}) = [-26,-18]$. All other cosmological parameters in the models, such as the amplitude of the power spectrum, were fixed to their Planck 2018 CMB values. 

The likelihood we chose for the UV luminosity function is that of the so-called Model I in~\cite{Sabti_2022a}, with its corresponding formalities and data. In the case of EFTofLSST, we selected the BOSS and eBOSS data sets as in\cite{Simon:2022csv}. It must be noted that some assumptions in the likelihoods have been made under the CDM paradigm only and would need to be amended for the case of SFDM. Taking into account these caveats, the results reported in the following may be stronger than in the case in which some of the assumptions are corrected for SFDM.

\begin{figure}[htp!]
\includegraphics[width=\columnwidth]{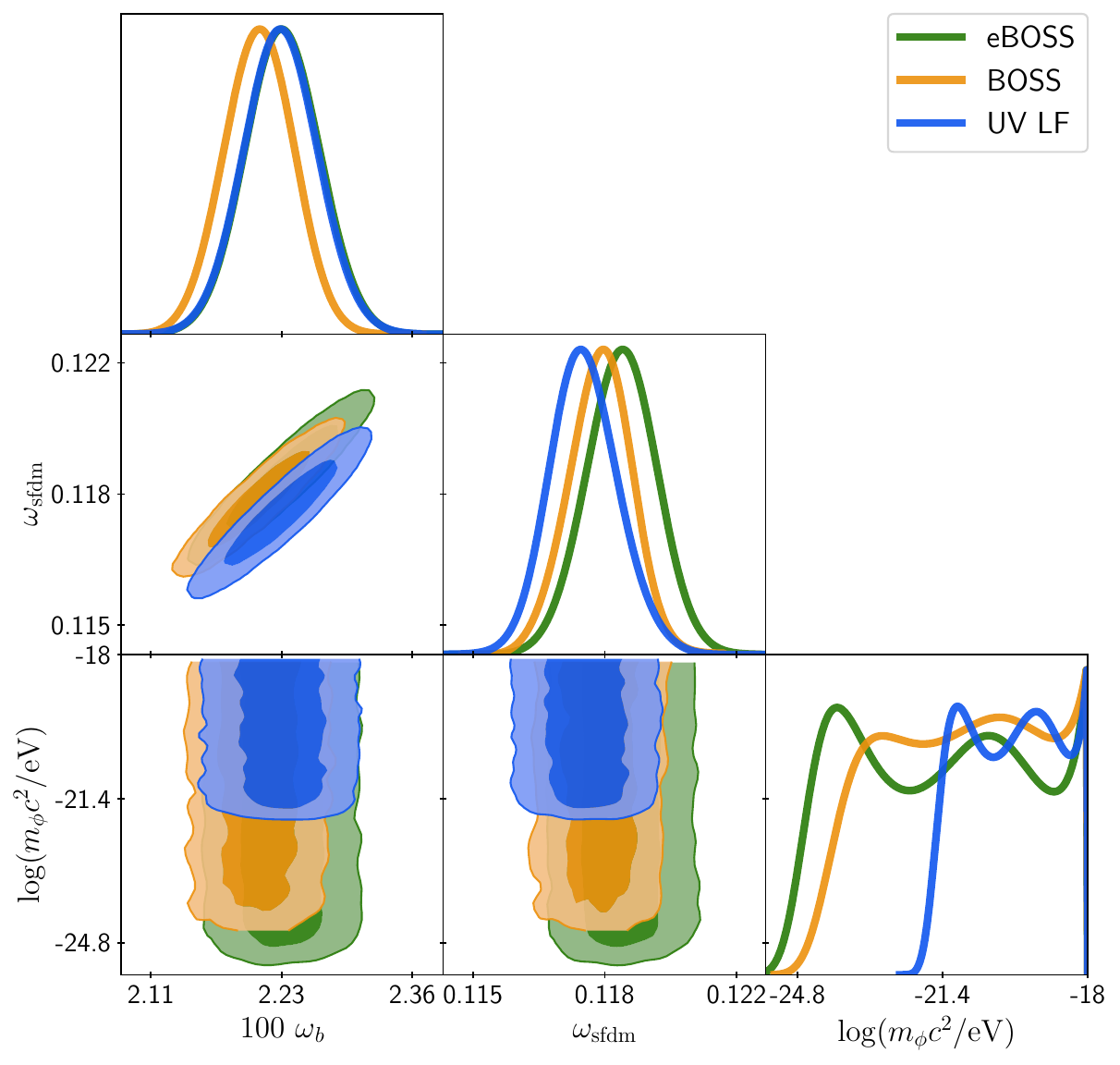}
\includegraphics[width=\columnwidth]{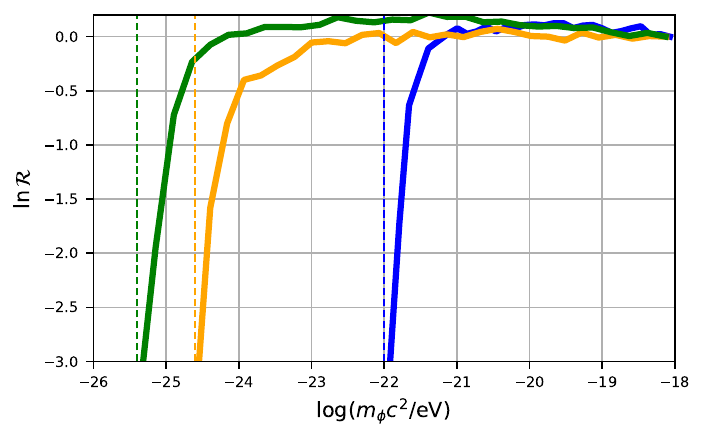}
\caption{\label{fig:Gallumi} (Top panel) Triangle plots obtained from the MCMC code \texttt{MontePython} using the likelihood of the UV galaxy luminous function, for parameters $\omega_b$, $\omega_{{\rm{sfdm}}}$ and $\log (m_\phi c^2/\mathrm{eV})$. (Bottom panel) Plot of the shape distortion function $\mathcal{R}$ obtained from the posterior distribution obtained for the field mass $m_\phi$. See the text for more details.}
\end{figure}

The resulting posterior distributions for the physical densities of baryons $\omega_b$ and SFDM $\omega_{{\rm{sfdm}}}$, and the field mass $m_\phi$, after marginalizing over the nuisance parameters of the likelihoods, are shown in Fig.~\ref{fig:Gallumi}. As expected, the separate constraints on the physical densities of baryons and SFDM are practically the same, as they are influenced mostly by the previously assumed priors.

Not surprisingly, the posterior distribution of the field mass shows that each data set only constrains $m_\phi$ from below, which means that the likelihoods are insensitive to variations of $m_\phi$ above a certain threshold value. The data set with the most constraining power is the UV luminosity function, whereas eBOSS is the less constraining one. 

To properly calculate the lower bounds for $m_\phi$ from the confidence regions in Fig.~\ref{fig:Gallumi}, we use the method in~\cite{astone1999inferring,dagostini2000confidence,Gariazzo:2019xhx} to obtain prior independent constraints by means of the so-called shape distortion function $\mathcal{R}$, appropriate for the so-called open likelihoods as in the present case. As explained in~\cite{Gariazzo:2019xhx}, the function $\mathcal{R}$ will allow us to use the data to define the region below which $m_\phi$ is disfavored, regardless of the prior assumptions we have chosen.

An advantage of the function $\mathcal{R}$ is that for its calculation we only need to know the posterior distribution of the field mass $m_\phi$, which we obtained from the code \texttt{MontePython}. The resultant shape distortion function is shown in the lower panel of Fig.~\ref{fig:Gallumi} on the logarithmic scale. Note that $\mathcal{R} \to 1$ for large values of the field mass, in this case $\log (m_\phi c^2/\mathrm{eV}) \to -18$ which is the upper value in our prior range. The sharp decay of $\mathcal{R}$ at lower values of $m_\phi$ helps us to calculate an appropriate lower bound. Following the convention in~\cite{astone1999inferring,dagostini2000confidence,Gariazzo:2019xhx}, it can be seen that if $\ln \mathcal{R} = -3$ (moderate level according to Jeffrey's scale), we can say that the data strongly favor the regions $\log (m_\phi c^2/\mathrm{eV}) > -25.4$ for eBOSS, $\log (m_\phi c^2/\mathrm{eV}) > -24.6$ for BOSS, and $\log (m_\phi c^2/\mathrm{eV}) > -22$ for UV LF\footnote{If we follow the standard wisdom and calculate the lower bound at the $95\%$ confidence level of the histograms, we find instead that $\log (m_\phi c^2/\mathrm{eV}) > -24.5$ for eBOSS, $\log (m_\phi c^2/\mathrm{eV}) > -23.8$ for BOSS, and $\log (m_\phi c^2/\mathrm{eV}) > -21.4$ for UV LF.}.

\section{Comparison with the original field variables $\left(\phi\, , \dot{\phi}\right)$}\label{comparison}
This section is dedicated to the comparison of the numerical solutions obtained from our polar method to those of the original scalar field equations of motion. To do this, we use the same Boltzmann code \texttt{CLASS} to provide the numerical results, so that they are subject to the same numerical methods and limitations of the code in the two cases. In the amended version, the polar variables are solved as a separate dark matter component, while the field variables are solved using the scalar field equations of the quintessence module already implemented in \texttt{CLASS}.

\subsection{Background quantities}
The first comparison of background quantities is for the variables $\phi$ and $\dot{\phi}$, which are the dynamical ones in the KG equation~\eqref{eq:2d} and the field potential. The relationship between the original and polar variables can be found from the transformation equations~\eqref{eq:dynsyscart}, in the form of
\begin{equation}
    \kappa \phi = - \frac{2 \sqrt{6}}{y_1} e^\beta \cos (\theta/2) \, , \quad \frac{\kappa \dot{\phi}}{m_{\phi}}  = \frac{2 \sqrt{6}}{y_1} e^\beta \sin (\theta/2) \, . \label{eq:phi-phiprime}
\end{equation}

We solved the KG equation~\eqref{eq:2d} separately, but with the same initial conditions as in the polar case through transformation~\eqref{eq:phi-phiprime}. It is not possible to accurately follow the numerical evolution after the onset of the field oscillations, and then we only solved the KG equation~\eqref{eq:2d} up to the equivalent time to $\theta_\star = 100\pi$ ($2m_{\phi} t_\star \simeq 100\pi$). After this time, the equations of motion are set directly to $\dot{\phi} =0$ and $\ddot{\phi} =0$, which means that the late-time solutions of the field variables are just $\phi (t > t_\star) = \phi(t_\star)$ and $\dot{\phi} (t > t_\star) = \dot{\phi} (t_\star)$ (and the density remains artificially constant afterwards).

\begin{figure}[htp!]
\includegraphics[width=\columnwidth]{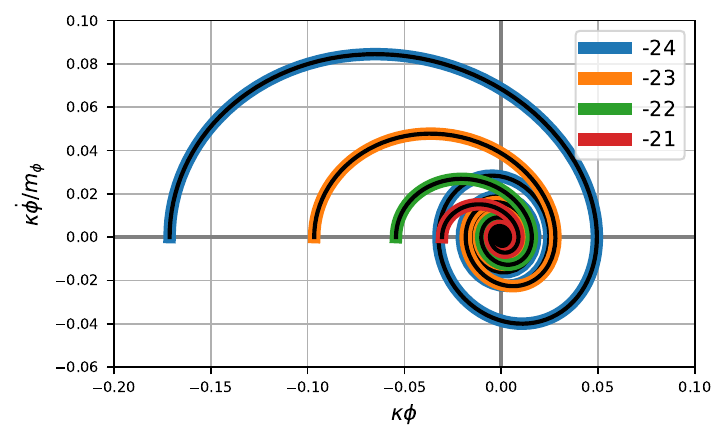}
\caption{\label{fig:phase-space} The phase space of the field variables $\phi$ and $\dot{\phi}$, written in their dimensionless form as in Eqs.~\eqref{eq:phi-phiprime}. The colored curves are numerical solutions from the original field variables, whereas the black curves are those obtained from the polar method. The curve labels represent the value of the field mass $m_{\phi}$ in units of $\mathrm{eV}$. See the text for more details.}
\end{figure}

The two sets of solutions, the original and the polar ones, are plotted in compact form in the phase space shown in Fig.~\ref{fig:phase-space}. The thick curves correspond to the system $\phi-\dot{\phi}$, with the different colors representing the field mass $m_{\phi}$, while the solutions of the polar system, all in black lines, are superimposed. We see that the agreement between the corresponding curves is exact up to $t=t_\star$.

\begin{figure*}[htp!]
\includegraphics[width=\textwidth]{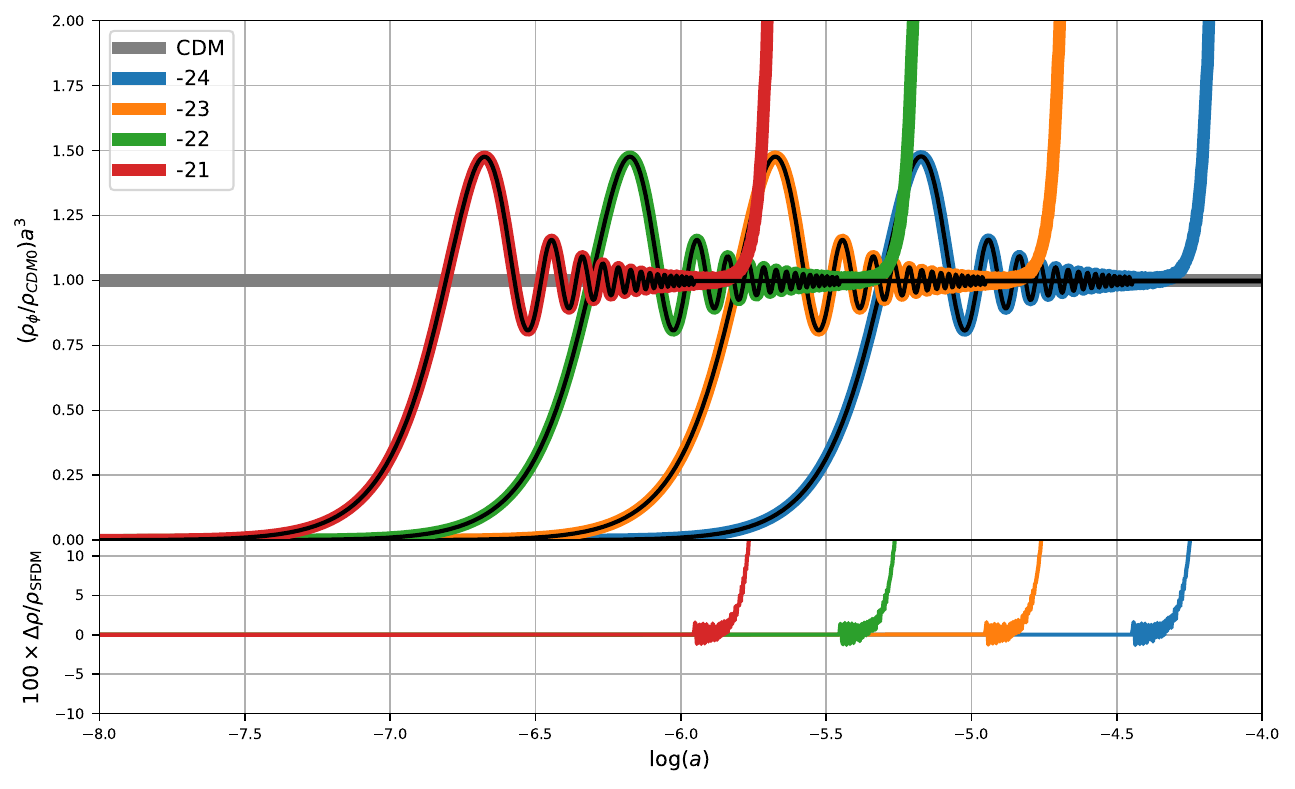}
\caption{\label{fig:sfdm-scf}The evolution of the field density $\rho_\phi$ for the same cases as in Fig.~\ref{fig:rho}. (Top panel) Colored curves represent the numerical solution of the original field variables $\phi$ and $\dot{\phi}$, while the corresponding solutions from the polar method are superimposed (black curves). The divergent behavior of the colored curves is due to the cutoff applied to the oscillations of the field solutions at $\theta_\star = 60\pi$. (Bottom panel) The relative difference between the curves in the upper panel for the same value of the field mass $m_{\phi}$. The difference is close to zero except for the last part, once the oscillations of the field solutions are cut off. See the text for more details.}
\end{figure*}

The same comparison exercise for field density $\rho_\phi$ is shown in Fig.~\ref{fig:sfdm-scf}, using the same colors for the different curves as in Fig.~\ref{fig:phase-space}. Furthermore, we normalize the density to the present value of the CDM density $\rho_{CDM 0}$, to highlight that the final value of the field densities coincides with the equivalent CDM case. The upper panel shows that the two sets of solutions coincide exactly, including the oscillatory phase, which is also confirmed by the comparison in the lower panel: the discrepancies appear at late times in the oscillatory phase and the cutoff in the solution of the field variables (see the explanation below Eq.~\eqref{eq:phi-phiprime}).

\begin{figure}[htp!]
\includegraphics[width=\columnwidth]{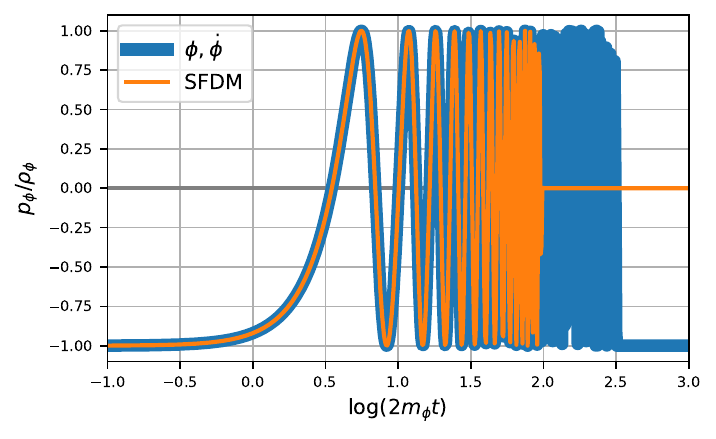}
\caption{\label{fig:EOS-phi} The numerical solution of the scalar field EOS $w_\phi$, calculated from the pressure to density ratio, as a function of the variable $2m_{\phi} t$, under which all cases collapse with different field masses $m_{\phi}$ collapse into a single curve. The blue curve is the solution of the original variables $\phi$-$\dot{\phi}$ (note that $w_\phi =-1$ after the oscillations are cut off is just an artifact), whereas the orange curve corresponds to the polar method. See the text for more details.}
\end{figure}

As a final example, we show in Fig.~\ref{fig:EOS-phi} the SFDM EOS calculated directly from the pressure-to-density ratio $w_\phi = p_\phi/\rho_\phi$, using the same unit system as in Fig.~\ref{fig:EOS}. The variable in the horizontal axis is the dimensionless quantity $2m_{\phi} t$, under which all curves corresponding to a given field mass $m_{\phi}$ become the same curve. The EOS oscillates rapidly around the zero value, and we again see that there is excellent agreement between the numerical results of the two approaches.

\subsection{Linear density perturbations}
We repeated the comparison of the solutions with the case of linear density perturbations. This time we solved the linearly perturbed KG equation~\eqref{eq:13}, again using the same initial conditions for the two sets of variables, the originals $\varphi$ and $\dot{\varphi}$ and the polar ones $\delta_0$ and $\delta_1$. Although the perturbed polar variables are $\alpha$ and $\vartheta$, see Eqs.~\eqref{eq:22a}, recall that our final perturbed variables are those of Eqs.~\eqref{eq:d0d1} and their corresponding equations of motion~\eqref{eq:newdeltas}.

The transformation from the polar variables to the field ones is given by the expressions
\begin{subequations}
\label{eq:deltas-to-varphi}
\begin{eqnarray}
    \kappa \varphi &=& -\frac{\sqrt{6}}{y_1} e^\beta \left[ \delta_0 \cos (\theta/2) - \delta_1 \sin (\theta/2) \right] \, , \\
    \frac{\kappa \dot{\varphi}}{m_{\phi}} &=& \frac{\sqrt{6}}{y_1} e^\beta \left[ \delta_0 \sin (\theta/2) + \delta_1 \cos (\theta/2) \right] \, .
\end{eqnarray}
\end{subequations}
We only used Eqs.~\eqref{eq:deltas-to-varphi} to set the initial conditions of the field variables $\varphi$ and $\dot{\varphi}$ in correspondence with those of the polar variables, and then the field equation~\eqref{eq:13} was solved separately.

\begin{figure*}[htp!]
\includegraphics[width=\columnwidth]{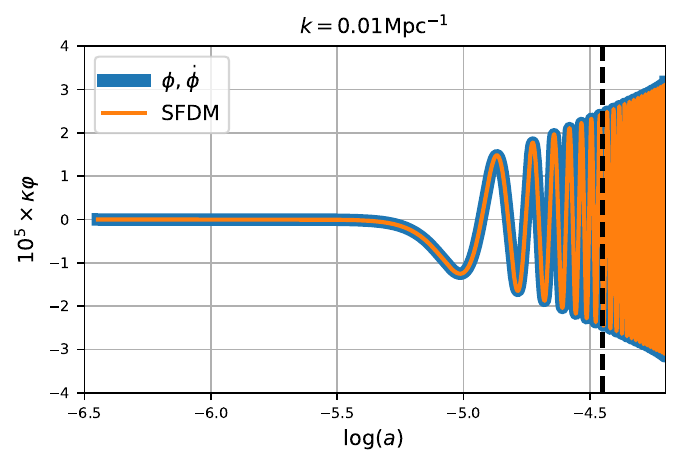}
\includegraphics[width=\columnwidth]{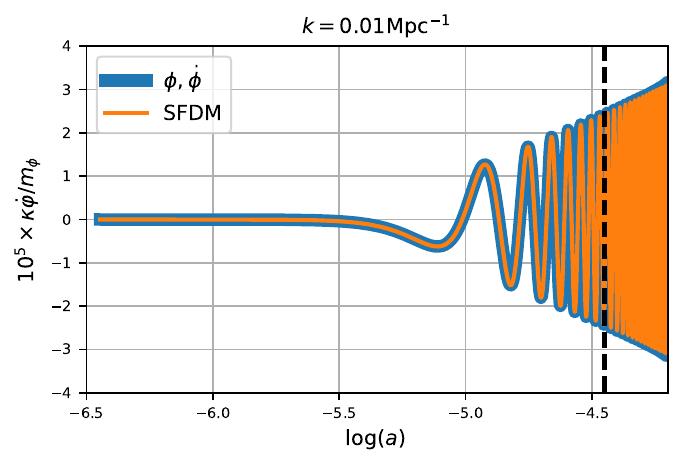}
\includegraphics[width=\columnwidth]{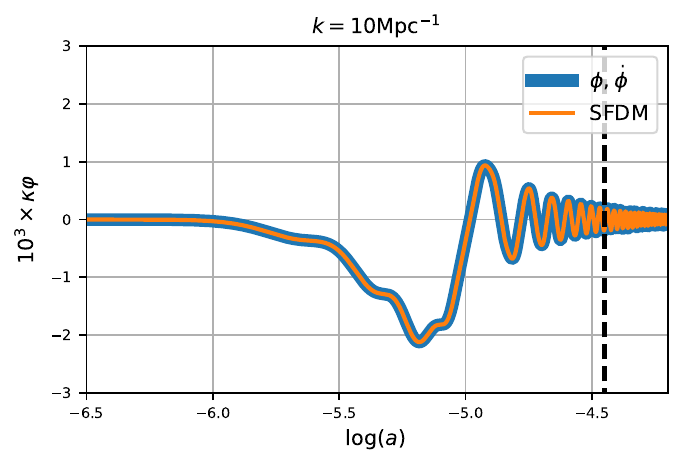}
\includegraphics[width=\columnwidth]{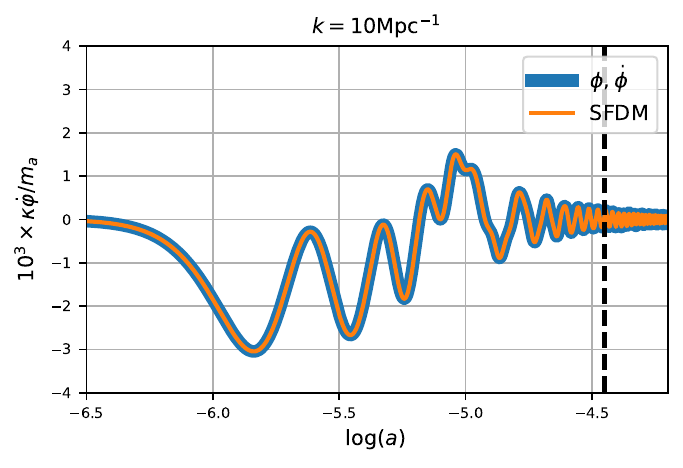}
\caption{\label{fig:varphi_sfdm_scf} The evolution of the perturbation variables $\varphi$ (left column) and $\dot{\varphi}$ (right column), as obtained from both the original field variables via Eq.~\eqref{eq:13} and the polar method in Eqs.~\eqref{eq:newdeltas}, see also Eqs.~\eqref{eq:deltas-to-varphi}. The upper panels show the solutions for large scales, whereas the lower panels show them for small scales.  It can be seen that the curves of the two methods agree completely even beyond the cutoff time for the polar variables at $\theta_\star =30\pi$ (indicated by the black-dashed vertical line). See the text for more details.}
\end{figure*}

We then show in Fig.~\ref{fig:varphi_sfdm_scf} the evolution of the perturbation variables $\varphi$ and $\dot{\varphi}$ as a function of the scale factor and for two values of the wavenumber $k$: $0.01 \, \mathrm{Mpc}^{-1}$ for large scales (top panels) and $10 \, \mathrm{Mpc}^{-1}$ for small scales (bottom panels).

In the two cases, the solutions from the polar variables via Eqs.~\eqref{eq:deltas-to-varphi} are superimposed on those of the field variables (obtained directly from Eq.~\eqref{eq:13}), which are identical up to the time the numerical solutions were followed ($\theta_\star =30\pi$ for the polar variables and $\theta_\star = 100\pi$ for the field ones). Note that the agreement goes beyond the cutoff point of the polar variables, which means that the cutoff of the trigonometric functions~\eqref{eq:trig-cutoff} delivers from Eqs.~\eqref{eq:newdeltas} the expected results of the original field variables.

\section{Discussion}\label{discussion}

Among the diverse theoretical proposals to describe dark matter, SFDM constitutes a compelling candidate to play the role of the CDM component of the universe. The dynamics of such a particle is modeled through a scalar field endowed with a scalar field potential. In this work, we were particularly interested in the free case of a real scalar field. Given the oscillatory nature of the SFDM when it behaves as CDM, we must be careful to properly handle the differential equations for both background and linear perturbations, in such a way that numerically they were easily solved, allowing us to keep track of the evolution of the scalar field, and of all physical quantities built from it. By posing the Klein-Gordon equations (background and linear perturbations) as a dynamical system once new variables are introduced, we were able to describe the evolution of the SFDM as a system of first-order differential equations. 

With this prescription, instead of solving for the original scalar field variables $\left(\phi\, , \dot{\phi}\, ;\varphi\, , \dot{\varphi}\right)$, we solved a new set of dynamical variables: the polar variable $\theta$, the scalar field energy density $\beta$, and the perturbations $\delta_0\, , \delta_1$. We have also added the variable $y_1$ which is proportional to the ratio of the mass of the scalar field and the Hubble parameter. It has been a standard procedure in the literature to deal with the rapid oscillations of the SFDM by averaging the oscillating functions in a Hubble time, and then writing down a new set of averaged equations of motion that resemble a standard cosmological fluid. Within our approach, this new set of equations is easier to solve numerically and to include in a standard Boltzmann code. Therefore, there was no need to invoke any approximation or average to cancel out the rapid oscillations of the scalar field. The only consideration of this kind was the introduction of the truncated trigonometric functions that we used (see Eq.~\eqref{eq:trig-cutoff} in Section~\ref{sec:III-A} ). In addition, the average procedure introduces an undesirable mismatch between the early- and late-time solutions and leaves unanswered the question of the sound speed of the density perturbations of the averaged fluid.

Our method does not require a separate evolution of the SFDM equations of motion, but just a straightforward transformation of the original field equations. Our transformed system of equations remains the same throughout the full evolution, and this applies both for the background and for the linear density perturbations. Moreover, in the case of linear perturbations, we do not need to define an explicit expression for the sound speed, which is a slippery quantity in the fluid approximation. Nevertheless, in Appendices~\ref{sec:fld_approx} and~\ref{sec:fluid-perts} we show the complete equivalence between our equations and the fluid ones, which only required the implicit definition of the sound speed. Our approach also deals successfully with the rapid oscillations in comparison with the original field variables. 

The SFDM formulation we present in this work gives accurate predictions on observables such as the CMB anisotropies and the MPS. Moreover, it is possible to put constraints on the main parameter of the model, the SFDM mass $m_{\phi}$. This is of great relevance because, on the one hand, we need numerical solutions with a high precision level to put constraints on the SFDM model in light of new data from present and future surveys. On the other hand, the cosmological dynamics of the SFDM from the big bang until the linear perturbation regime needs to be properly solved in order to set the initial conditions for numerical simulations of the nonlinear process of structure formation. In this sense, our approach offers a mathematical and numerical treatment for the SFDM model, which is advantageous in that it suitably solves the cosmological dynamics of such scalar field coupled to the Einstein-Boltzmann system.

\begin{acknowledgments}
FXLC acknowledges Beca CONACYT and FORDECYT-PRONACES-CONACYT for support of the present research under Grant No. CF-MG-2558591. This work was partially supported by Programa para el Desarrollo Profesional Docente; Direcci\'on de Apoyo a la Investigaci\'on y al Posgrado, Universidad de Guanajuato; CONACyT M\'exico under Grants No. A1-S-17899, No. 286897, No. 297771, No. 304001; and the Instituto Avanzado de Cosmolog\'ia Collaboration. We acknowledge the use of open-source software: Python~\cite{van1995python,Hunter:2007}, numpy~\cite{van_der_Walt_2011}, scipy~\cite{Virtanen:2019joe}, astropy~\cite{Astropy:2018wqo}.
\end{acknowledgments}

\appendix

\section{The fluid approximation \label{sec:fld_approx}}
Here we discuss the possibility of rewriting the equations of motion~\eqref{eq:newdeltas} for the linear density perturbations in their fluid counterparts. Using Eq.~\eqref{eq:old-to-new-deltas-b}, we find that Eq.~\eqref{eq:newdeltas-a} becomes
\begin{equation}
    \delta^\prime_0 =  - \frac{1}{aH} (1+w_\phi) \Theta_\phi -3\sin\theta \delta_1  - (1+w_\phi) \frac{\bar{h}^\prime}{2} \, , \label{eq:deltarho_f0}
\end{equation}
where we have also used the relation~\eqref{eq:eos-Ophi-b}. From the definition of the adiabatic sound speed $c^2_s = \delta p_\phi/\delta \rho_\phi$ we find that
\begin{equation}
    c^2_s \delta_0 = \sin \theta \, \delta_1 - \cos \theta \, \delta_0 \, . \label{eq:deltarho_f0a}
\end{equation}
The preceding equation can be used to substitute the second term on the right-hand side of Eq.~\eqref{eq:deltarho_f0}. After some manipulations, we get the following result,
\begin{subequations}
\label{eq:deltarho_f}
\begin{equation}
    \delta^\prime_0 =  - (1+w_\phi) \Theta_\phi -3 \mathcal{H} (c^2_s - w_\phi) \delta_0  - (1+w_\phi) \frac{\bar{h}^\prime}{2} \, , \label{eq:deltarho_f1}
\end{equation}
where now the primes denote the derivatives with respect to the conformal time and $\mathcal{H}$ is the corresponding Hubble parameter. These calculations show that Eq.~\eqref{eq:newdeltas-a} is completely equivalent to the equation of motion for the density contrast in the fluid approximation.

Similarly, we can combine Eqs.~\eqref{eq:dy_sys}, \eqref{eq:newdeltas} and~\eqref{eq:deltarho_f0a} to write an equation for the momentum density. After some lengthy but otherwise straightforward manipulations, we find that
\begin{equation}
    \left[(1+w_\phi) \Theta_\phi \right]^\prime = \mathcal{H} (3w_\phi -1) (1+w_\phi) \Theta_\phi + k^2 c^2_s \delta_0 \, . \label{eq:deltarho_f2}
\end{equation}
\end{subequations}
Equations~\eqref{eq:deltarho_f} are the standard fluid equations for linear density perturbations and then show that the polar equations~\eqref{eq:newdeltas} are their faithful representation.

\section{Linear density perturbations after the cutoff point \label{sec:fluid-perts}}
It is illustrative to write Eqs.~\eqref{eq:newdeltas} after the cutoff on the rapid field oscillations. For times $t > t_\star$ we find
\begin{equation}
    \left( 
    \begin{array}{c}
    \delta_0  \\
    \delta_1 
    \end{array}
    \right)^\prime \simeq \frac{k^2}{k^2_J} \left( 
    \begin{array}{cc}
        0 & -1 \\
        1 & 0
    \end{array}
    \right) - \left( 
    \begin{array}{c}
    \bar{h}^\prime/2 \\
    0 
    \end{array}
    \right) \, . \label{eq:delta_harmonic}
\end{equation}
where again a prime denotes derivatives with respect to $N = \ln a$.

The system~\eqref{eq:delta_harmonic} has the same structure as a forced harmonic oscillator with frequency $\omega = k^2/k^2_J$. As explained in~\cite{Urena-Lopez:2015gur}, if we assume that $\omega$ is a constant (as is the case during RD), for large scales, the solution is the same as that of CDM: $\delta_0 \simeq - \bar{h}/2$ and $\delta_1 \simeq \mathrm{const.}$ For small scales, the solutions are oscillatory and given by combinations of functions $\cos(\omega N)$ and $\sin(\omega N)$. These behaviors agree with the full numerical results in Fig.~\ref{fig:deltas}.

To obtain the corresponding fluid equations, we first notice that the cutoff method applied to Eqs.~\eqref{eq:13} and~\eqref{eq:deltarho_f0a} results in the following equations,
\begin{equation}
    \Theta_\phi \simeq \frac{k^2}{a H y_1} \delta_1 \, , \quad  (c^2_s -w_\phi) \delta_0 \simeq 0 \, . \label{eq:fluid-nonrel}
\end{equation}
The first expression in Eqs.~\eqref{eq:fluid-nonrel} reveals a direct relation between the divergence of the fluid velocity and variable $\delta_1$, while the second suggests a non-varying value of the fluid's EOS (that is, $d w_\phi/d\rho_\phi =0$, see~\cite{Ma:1995ey}).

Repeating the same procedure above for the calculation of Eqs.~\eqref{eq:deltarho_f}, we find their counterpart for times $t > t_\star$,
\begin{equation}
    \delta^\prime_0 \simeq  - \Theta_\phi - \frac{\bar{h}^\prime}{2} \, , \quad \Theta_\phi^\prime \simeq - \mathcal{H} \Theta_\phi + \frac{k^2}{4 m_{\phi}^2 a^2} \delta_0 \, , \label{eq:fluid_nonrel1}
\end{equation}
where a prime now denotes derivatives with respect to conformal time. Finally, the second of Eqs.~\eqref{eq:fluid_nonrel1} suggests a scale-dependent sound speed given by
\begin{equation}
    c^2_s \simeq \frac{k^2}{4m_{\phi}^2 a^2}\, .
    \label{cs2}
\end{equation}

One note on the sound speed of the perturbations is taken in turn. The discussion above gives a clear answer for the sound speed after the cutoff in the field oscillations, but there is none for the right expression of $c^2_s$ before that. Our formalism for the perturbations does not require an explicit expression of $c^2_s$, but the latter is necessary for a correct fluid formulation of the SFDM linear perturbations.

A general expression of $c^2_s$ was found in~\cite{Hwang:2009js} in the comoving gauge of the scalar field, but the same authors argue in~\cite{Hwang:2021vuq} that such expression cannot be transformed into the synchronous gauge, under which SFDM lacks a definite functional form of $c^2_s$. Despite this, the expression for $c^2_s$ of~\cite{Hwang:2009js} has been extensively used in the literature for the fluid approximation of SFDM in the synchronous gauge. 

An example is the so-called Effective Fluid Approximation (EFA) discussed in~\cite{Cookmeyer:2019rna}. The equations of motion of the EFA are
\begin{subequations}
\begin{eqnarray}
    \delta^\prime_0 &\simeq&  - \Theta_\phi - 3\mathcal{H} \langle c^2_s \rangle \delta_0 - \frac{\bar{h}^\prime}{2} \, , \label{eq:efa-a} \\ \Theta_\phi^\prime &\simeq& - \mathcal{H} \Theta_\phi + k^2 \langle c^2_s \rangle \delta_0 \, , \label{eq:efa-b}
\end{eqnarray}
where the cycle-averaged sound speed is 
\begin{equation}
    \langle c^2_s \rangle = \frac{k^2/(4m_{\phi}^2 a^2)}{1+k^2/(4m_{\phi}^2 a^2)} \, . \label{eq:efa-c}
\end{equation}
\end{subequations}

It can be seen that our Eqs.~\eqref{eq:fluid_nonrel1} can be obtained from the EFA if the sound speed is given by the so-called non-relativistic expression $\langle c^2_s \rangle \simeq k^2/(4m_{\phi}^2 a^2)$. However, the EFA considers an extra term in the equation of motion of $\delta_0$, see Eq.~\eqref{eq:efa-a}.

Noticing this discrepancy, a comparison was made between the fluid approximation and our method in~\cite{Cookmeyer:2019rna}, see their Appendix~B, and no relevant differences were found in the numerical results. To understand this, we use Eqs.~\eqref{eq:fluid-nonrel} and write the EFA equations~\eqref{eq:efa-a} and~\eqref{eq:efa-b} in terms of our variables $\delta_0,\delta_1$ and $N=\ln a$. We find that
\begin{equation}
    \left( 
    \begin{array}{c}
    \delta_0  \\
    \delta_1 
    \end{array}
    \right)^\prime \simeq \frac{k^2}{k^2_J} \left( 
    \begin{array}{cc}
        -3/y_1 & -1 \\
        1 & 0
    \end{array}
    \right) - \left( 
    \begin{array}{c}
    \bar{h}^\prime/2 \\
    0 
    \end{array}
    \right) \, . \label{eq:efa_harmonic}
\end{equation}

We now compare Eqs.~\eqref{eq:efa_harmonic} with Eqs.~\eqref{eq:delta_harmonic}, and spot the extra term $3/y_1$. The latter is very small, given that $y_1 = 2m_{\phi}/H$ and then $y_1 \gg 1$ after the start of rapid field oscillations. This is the reason why the comparison in~\cite{Cookmeyer:2019rna} did not find differences between our approach and the EFA. It can be concluded that our method correctly picks up the only terms in the equations of motion that are valid in the limit $H/m_{\phi} \ll 1$.~\footnote{The study in~\cite{Cookmeyer:2019rna}, and others in which the EFA is used, missed to neglect the extra term in Eq.~\eqref{eq:efa-a} because they did not analyze the joint term $\mathcal{H}\langle c^2_s \rangle$, otherwise they would have realized that it is of order $\mathcal{O}(H/m_{\phi})$. The same applies to the EFA used in~\cite{Hlozek:2014lca} that includes additional terms from the gauge transformation: These terms are also of the order $\mathcal{O}(H/m_{\phi})$ and can be neglected. This effectively leads the EFA to become again Eqs.~\eqref{eq:fluid_nonrel1}.}

\section{Numerical implementation in \texttt{CLASS} \label{sec:class-details}}
Following the design of \texttt{CLASS}, we introduced a new module for the SFDM equations of motion, replicating the same structure as for other dark matter components. In this form, the contribution of the SFDM component could be called with its own parameters in any given parameter file of \texttt{CLASS} (for example, \textit{explanatory.ini}).

The equations of motion for the background variables~\eqref{eq:13} were included in the file \textit{background.c}, while those of the density perturbations~\eqref{eq:newdeltas} were included in the file \textit{perturbations.c}. Similarly, a shooting routine was incorporated into the file \textit{input.ini} to adjust the initial conditions of the dynamical variables. In each case, the SFDM quantities were added to the matter budget so that they contribute correctly to the right-hand side of the Einstein equations, for both the background and the linearly perturbed ones.

Given the oscillatory nature of all SFDM quantities, we applied the cutoff procedure to the sine and cosine functions that appear in their definition, see Eqs.~\eqref{eq:trig-cutoff}. For example, the background pressure was written as $p_\phi = -\cos_\star \theta \cdot \rho_\phi$, and then effectively $p_\phi =0$ for $t > t_\star$. Another case would be the combination $p_\phi + \rho_\phi = (1-\cos_\star \theta) \rho_\phi$, so that $p_\phi + \rho_\phi = \rho_\phi$ for $t > t_\star$.

The same was applied to the perturbed quantities, as in the density perturbation written in the form $\delta \rho_\phi = \rho_\phi \cdot \delta_0$, which is required for the right-hand side of the perturbed Einstein equations. Similarly, the momentum density perturbation $(p_\phi + \rho_\phi) \Theta_\phi$ was written as in Eq.~\eqref{eq:old-to-new-deltas-b}, that is, 
\begin{subequations}
\label{eq:momentum-pert}
\begin{equation}
    (\rho_\phi +p_\phi) \Theta_\phi = \frac{k^2 \rho_\phi}{a H y_1} \left[ (1 -\cos_\star \theta) \delta_1 - \sin_\star \theta \, \delta_0 \right] \, . \label{eq:momentum-pert-a}
\end{equation}
The result for late times $t > t_\star$ is well defined and is given by
\begin{equation}
    (\rho_\phi +p_\phi) \Theta_\phi = \frac{k^2 \rho_\phi}{a H y_1} \delta_1 \, . \label{eq:momentum-pert-b}
\end{equation}
\end{subequations}

It should be noted that in some parts of the file \textit{perturbations.c} one requires to calculate the velocity divergence $\Theta$, which is one of the fundamental variables in the fluid formalism for linear perturbations of density and momentum. However, the calculation of $\Theta_\phi$ itself is problematic for SFDM, as can be clearly seen in Eq.~\eqref{eq:momentum-pert-a}: One needs to divide by the quantity $(p_\phi + \rho_\phi)$, which passes through zero during the phase of rapid field oscillations. As explained above, the calculation of $\Theta_\phi$ is not needed in the perturbed Einstein equations, and then we only used Eq.~\eqref{eq:momentum-pert-b} as an additional source of the perturbed momentum density.

A technical aspect not always mentioned for SFDM models and its Boltzmann code implementation is that it is always required a small quantity of CDM component. This is because in order to be consistent with the synchronous gauge, the Einstein field equations have to be solved in the comoving frame of the CDM fluid, that is, $\Theta_{cdm} = 0$. In fact, an automatic feature in the current versions of \texttt{CLASS} and its amended versions for SFDM, $\Omega_{cdm} = 10^{-10}$ once the amount of CDM is set to zero in the input file \textit{explanatory.ini}. This ensures that there will be practically a null contribution of CDM, so that SFDM is the dominant non-relativistic matter component, and also helps us to avoid the known problem that the synchronous gauge is not completely fixed for density perturbations when the SFDM is the only DM component~\cite{Hwang:2021vuq}.

One last note is that the evolution of the perturbations should start well before the start of the rapid oscillations. We include an additional condition in the file \textit{perturbations.c} to guarantee that the calculations begin only if $m_{\phi}/H < 0.01$. This gives the system~\eqref{eq:newdeltas} enough time to reach its attractor solution starting from the initial conditions~\eqref{eq:delta_ini}.

\nocite{*}

\bibliography{biblio}

\end{document}